\documentclass[aps,superscriptaddress]{revtex4}
\usepackage{amssymb,amsmath,epsfig}
\usepackage{subfigure}
\usepackage{multirow}
\usepackage[colorlinks=true, pdfstartview=FitV, linkcolor=blue, citecolor=red, urlcolor=magenta, breaklinks=true]{hyperref}
\begin{document}
\title{Quasinormal modes and shadow of a Schwarzschild black hole with GUP}

\author{M. A. Anacleto}\email{anacleto@df.ufcg.edu.br}
\affiliation{Departamento de F\'{\i}sica, Universidade Federal de Campina Grande
Caixa Postal 10071, 58429-900 Campina Grande, Para\'{\i}ba, Brazil}

\author{J. A. V. Campos}\email{joseandrecampos@gmail.com}
\affiliation{Departamento de F\'isica, Universidade Federal da Para\'iba, 
Caixa Postal 5008, 58051-970 Jo\~ao Pessoa, Para\'iba, Brazil}

\author{F. A. Brito}\email{fabrito@df.ufcg.edu.br}
\affiliation{Departamento de F\'{\i}sica, Universidade Federal de Campina Grande
Caixa Postal 10071, 58429-900 Campina Grande, Para\'{\i}ba, Brazil}
\affiliation{Departamento de F\'isica, Universidade Federal da Para\'iba, 
Caixa Postal 5008, 58051-970 Jo\~ao Pessoa, Para\'iba, Brazil}
 
\author{E. Passos}\email{passos@df.ufcg.edu.br}
\affiliation{Departamento de F\'{\i}sica, Universidade Federal de Campina Grande
Caixa Postal 10071, 58429-900 Campina Grande, Para\'{\i}ba, Brazil}

\begin{abstract}  
We consider quantum corrections for the Schwarzschild black hole metric by using the generalized uncertainty principle (GUP) to investigate quasinormal modes, shadow and their relationship in the eikonal limit. We calculate the quasinormal frequencies of the quantum-corrected Schwarzschild black hole by using the sixth-order Wentzel-Kramers-Brillouin (WKB) approximation, and also perform a numerical analysis that confirms the results obtained from this approach.
We also find that the shadow radius is nonzero even at very small mass limit for finite GUP parameter.

\end{abstract}

\maketitle
\pretolerance10000

\section{Introduction}
The direct detection of gravitational waves that arise from the merger of black holes and neutron stars has been presented by the LIGO-VIRGO collaboration~\cite{Abbott:2016blz,TheLIGOScientific:2017qsa}. 
A ringdown phase arises from the signal of the gravitational waves detected in the black hole merger process. Thus, its frequency spectrum and damping time can be analyzed by studying quasi-normal modes.
Quasinormal modes are complex frequencies that extract information about how black holes relax after the disturbance stops acting on them.
A well-known way of analyzing quasinormal modes is through the Wentzel-Kramers-Brillouin (WKB) approximation. 
The first works using the WKB approach to find quasinormal modes were done by Schutz 
and Will~\cite{Schutz:1985zz}.
Improvements in the approach were made using corrections up to second order~\cite{Iyer:1986np,Seidel:1989bp} 
and up to sixth order by Konoplya~\cite{Konoplya:2003ii}.
The quasinormal modes can also be determined numerically and developments for this procedure was described by Leaver~\cite{Leaver:1985ax,Leaver:1990zz} and also in more recent works~\cite{Konoplya:2011qq,Cardoso:2004fi} showing to be a method with good accuracy.

In General Relativity, black hole solutions have a central point in their spherical geometry, which is the existence of singularities.
However, these singularities lead to the breakdown of General Relativity and thus, in order to avoid such singularities, many studies have been proposed in order to obtain solutions of regular black holes through the deformation of Schwarzschild black holes. 
A regular black hole solution has been proposed initially by Bardeen in 1968~\cite{Bardeen} (for a review of regular black holes see~\cite{Ansoldi:2008jw}).
In~\cite{Kazakov:1993ha} Kazakov and Solodukhin has performed a deformation in the Schwarzschild solution, by introducing a quantum correction term, with the purpose of eliminating the singularity in the Schwarzschild metric.
An analysis of the quasinormal modes of the quantum-corrected Schwarzschild black hole has been investigated 
in~\cite{Konoplya:2019xmn}. And in~\cite{Hajebrahimi:2020xvo} the authors have explored the Hawking radiation from the quantum-corrected Schwarzschild black hole via the tunneling process. 
In addition, thermodynamic properties have been considered in~\cite{Kim:2012cma,Shahjalal:2019pqb,Lobo:2019put,Shahjalal:2019fig}.
In addition, a regular black hole solution can also be obtained by assuming a non-commutative spacetime. 
Quasinormal modes of noncommutative black holes have been studied by many authors~\cite{Giri:2006rc,Gupta:2015uga,Gupta:2017lwk,Ciric:2017rnf,Liang:2018uyk} and in~\cite{Anacleto:2019tdj} we have explored the absorption and scattering due to the noncommutative Schwarzschild black hole via Lorentzian smeared
mass distribution and in~\cite{Campos:2021sff} quasinormal modes and shadow radius were investigated for the noncommutative Schwarzschild black hole with Lorentzian distribution.

Also in an attempt to remove such singularities, a large number of works have considered astrophysical objects (effective radius greater than horizon)~\cite{Khodadi:2017eim}. Thus, we have for example 
the fuzzball paradigm~\cite{Mathur:2005zp}, gravastar~\cite{Mazur:2001fv,Beltracchi:2018ait} and the firewall~\cite{Braunstein:2009my,Almheiri:2012rt}.
In recent years, when considering the generalized uncertainty principle 
(GUP)~\cite{Das:2008kaa,Das:2009hs,Ali:2011fa,Buoninfante:2020cqz,Bosso1,Bosso2}, we have the formation of an object with an effective radius greater than the radius of the Schwarzschild horizon.
In addition, in~\cite{Anacleto:2020lel} we have considered the Schwarzschild black hole with quantum corrections implemented by the GUP and computed the absorption and the differential scattering cross section. We show that at zero mass limit the absorption does not cancel out. 
Recently, in~\cite{Anacleto:2020zhp} considering a self-dual black hole, we have investigated the scattering and absorption process and obtained a similar result. 
We can also mention the self-dual black hole~\cite{Modesto:2008im,Modesto:2009ve}, which is another example of a quantum-corrected black hole. 
In~\cite{Anacleto:2015mma}, by extending the thermodynamic analyzes performed in~\cite{Silva:2012mt}, the Hawking radiation from a self-dual black hole with GUP was investigated via tunneling formalism~\cite{Anacleto:2014apa,Anacleto:2015rlz,Anacleto:2015kca,Haldar:2018zyv,Haldar:2019fcz}.
{Moreover, the tunneling process of charged massive vector particles was explored in the context of noncommutative charged black holes in~\cite{Ovgun:2015box}.
Besides, an analysis of the quasinormal modes in the background of eternal-Alcubierre-warp-drive spacetime has been performed in~\cite{Jusufi:2017trn}.
Also, in~\cite{Gonzalez:2017zdz} the Hawking radiation and the quasinormal modes of a three-dimensional G\"odel black hole were analyzed by applying the tunneling formalism.  
In a study carried out in~\cite{Ovgun:2018gwt}, the authors have found exact solutions for the greybody factors and quasinormal modes of a massless scalar field in the 2+1 dimensional background of a cloud of strings black hole.}

Related to the investigation of quasinormal modes, the study of the shadow of black holes has been extensively explored in the literature~\cite{cunha2018shadows,mishra2019understanding,KONOPLYA20191,haroon2020shadow,bisnovatyi2018shadow,Khodadi:2021gbc,Zeng:2021dlj,CBambi,SunnyV,Vagnozzi,Vagnozzi2,Ovgun:2018tua,Ovgun:2020gjz,Ovgun:2021ttv}. This study is a very important tool to understand the properties of the black hole near the event horizon.
In addition, interest in studying the black hole shadow has increased due to advances and improvements in experimental techniques applied to the Event Horizon Telescope capturing the image of a supermassive black hole at the center of galaxy M87~\cite{event2019firstI,event2019firstVI}. 
To address these issues, it is also interesting the use of an alternative method for computing quasinormal modes at the eikonal limit that has been done by Mashhoon~\cite{mashhoon1985stability}.
In this regard, several works have been carried out to determine the quasinormal modes in the eikonal 
limit in order to relate the shadow radius to the real part of the quasinormal frequencies~\cite{stefanov2010connection,jusufi2020quasinormal,cuadros2020analytical,Moura:2021eln,Liu2021}.
{In~\cite{Okyay:2021nnh} the authors investigated some properties of a magnetically charged black hole, such as quasinormal modes, shadows, and greybody factors.}
Thus, in this work, we will also analyze the shadow radius and quasinormal modes at the eikonal limit to better understand the properties of the Schwarzschild black hole with quantum corrections near the event horizon.

The paper is organized as follows. In Sec.~\ref{qcmet} we will consider quantum corrections for a Schwarzschild black hole due to GUP, introduced 
in~\cite{Anacleto:2020lel}, in order to study quasinormal modes.
We will compute the quasinormal frequencies applying the WKB approximation method up to the sixth-order. We also perform a numerical analysis to confirm the WKB approach.
In addition, in Sec.~\ref{ng-sh} we investigate the shadow radius of the Schwarzschild black hole with quantum corrections due to GUP and its connection with quasinormal modes in the eikonal limit. The final comments are in Sec.~\ref{conclu}.

\section{Quasinormal frequencies}
\label{qcmet}
In this section, we will consider the Schwarzschild black hole with quantum corrections incorporated 
by the GUP~\cite{ADV, Tawfik:2014zca, KMM, Tawfik:2015kga, Gangopadhyay:2015zma} to investigate quasinormal modes. Thus, as shown recently in~\cite{Anacleto:2020lel}, the metric can be constructed by replacing the Schwarzschild horizon radius, $ r_h $, with the GUP radius, $ r_{hgup} $. 
{Below we briefly present some steps to incorporate quantum corrections into the Schwarzschild black hole metric.
Let us start with the GUP
\begin{eqnarray}
\label{gup}
\Delta x\Delta p\geq \frac{\hbar}{2}\left( 1-\frac{\alpha l_p}{\hbar} \Delta p 
+\frac{\beta l^2_p}{\hbar^2} (\Delta p)^2 \right),
\end{eqnarray}
where $\alpha$ and $ \beta $ are positive dimensionless parameters and $ l_p $ is the Planck length. 
Then adopting $ G=c=k_B=\hbar=l_p=1 $ and considering $ \Delta x\sim r_h $,
we obtain the following relationship
\begin{eqnarray}
\label{rdgup}
{\cal E}&\geq& E\left[1-\frac{4\alpha}{(\Delta x)}+ \frac{16\beta}{(\Delta x)^2}+\cdots \right]
= E\left[1-\frac{4\alpha}{r_h}+ \frac{16\beta}{r_h^2}+\cdots \right],
\end{eqnarray}
where $\mathcal{E} $ is identified with the GUP-corrected black hole energy. 
Now by assuming $E\sim M$, ${\cal E}\simeq M_{gup}$, $ r_h=2M $, 
the above relationship can be written in terms of mass as follows~\cite{Anacleto:2020lel}
\begin{eqnarray}
M_{gup}\geq M\left[1-\frac{4\alpha}{r_h}+ \frac{16\beta}{r_h^2}\right]= M\left( 1- \frac{2\alpha}{M}+\frac{4\beta}{M^2} \right).
\end{eqnarray}
Hence, for the radius of the horizon we have }
\begin{eqnarray}
{r}_{hgup} = {2M_{gup}}\geq r_h\left( 1- \frac{4\alpha}{r_h}+\frac{16\beta}{r^2_h} \right).
\end{eqnarray}
So we have the following line element:
\begin{eqnarray}
\label{metrsd}
ds^2={\cal F}(r) dt^2-{\cal F}(r)^{-1}dr^2-r^2\left(d\theta^2 + \sin^2\theta d\phi^2 \right),
\end{eqnarray}
where
\begin{eqnarray}
{\cal F}(r)=1-\frac{r_{hgup}}{r},
\end{eqnarray}
and $ {r}_{hgup} $ is given by 
\begin{eqnarray}
{r}_{hgup} = {2M_{gup}}=r_h\left( 1- \frac{2\alpha}{M}+\frac{4\beta}{M^2} \right).
\end{eqnarray}

Now, we apply the sixth-order WKB approximation method to compute quasinormal frequencies and compare with the result obtained numerically in order to check the accuracy of the WKB approach. Let us consider the case of the massless scalar field equation to describe the scattered wave in the background (\ref{metrsd}),  given by 
\begin{eqnarray}
\dfrac{1}{\sqrt{-g}}\partial_{\mu}\Big(\sqrt{-g}g^{\mu\nu}\partial_{\nu}\Psi\Big)=0 .
\end{eqnarray}
By applying a separation of variables into the equation above
\begin{eqnarray}
\Psi_{\omega l m}({\bf r},t)=\frac{{R}_{\omega l}(r)}{r}Y_{lm}(\theta,\phi)e^{-i\omega t},
\end{eqnarray}
where $Y_{lm}(\theta,\phi)  $ are the spherical harmonics and $ \omega $ is the frequency, then the radial equation for $ {R}_{\omega l}(r) $ is now written in the following form  
\begin{eqnarray}
\label{eqrad}
\dfrac{d^2{R}_{\omega l}(r_{\ast})}{dr^2} +\left[ \omega^2 -V_{eff} \right]{R}_{\omega l}(r_{\ast})=0,
\end{eqnarray}
being $ dr_{\ast}={\cal F}(r)^{-1}dr $ and $ V_{eff} $
 defined as the effective potential
\begin{eqnarray}
V_{eff}=\frac{{\cal F}(r)}{r}\frac{d{\cal F}(r)}{dr} + \frac{{\cal F}(r)l(l+1)}{r^2}.
\end{eqnarray}

\subsection{WKB Approach}
The formula for obtaining quasinormal frequencies up to the sixth order, found by Konoplya~\cite{Konoplya:2003ii}, is written as follows:
\begin{equation}
\dfrac{i\left(\omega_{n}^{2} - V_{0}\right)}{\sqrt{-2V_{0}''}} - \sum_{j=2}^{6}\Omega_{j}= n + \dfrac{1}{2},
\label{eqWKB}
\end{equation}
where $V_{0}$ is the maximum effective potential at point $r_{0}$. The indices $('')$ refer to the second derivative in relation to the tortoise coordinate and $\Omega_{j}$ are the corrections. 
We can get the point $r_{0}$ by solving the equation $V_{0}'=0$, i..e, 
\begin{eqnarray}
r_{0} & = & \frac{M}{2}\left(1 - \frac{2\alpha}{M} + \frac{4\beta}{M^2}\right)\Lambda(l)  
= \dfrac{M_{gup}}{2}\Lambda(l),
\end{eqnarray}
where
\begin{eqnarray}
\Lambda(l) & = & \dfrac{1}{l(l+1)}\left\lbrace 3\left[l(l+1)-1\right] + \sqrt{9 + l(l+1)\left[14 + 9l(l+1)\right]}\right\rbrace.
\end{eqnarray}
For $ M \rightarrow 0 $ we can expand the above equation in a  power series to obtain an expression that in this case is given by
\begin{equation}
r_{0} = \dfrac{2 \Lambda(l) \beta}{M} - \Lambda(l)\alpha + \frac{\Lambda(l)M}{2} + \mathcal{O}(M^{3}).
\end{equation}

\subsection{Scattering}
\begin{figure}[!htb]
 \centering
 \subfigure[]{\includegraphics[scale=0.30]{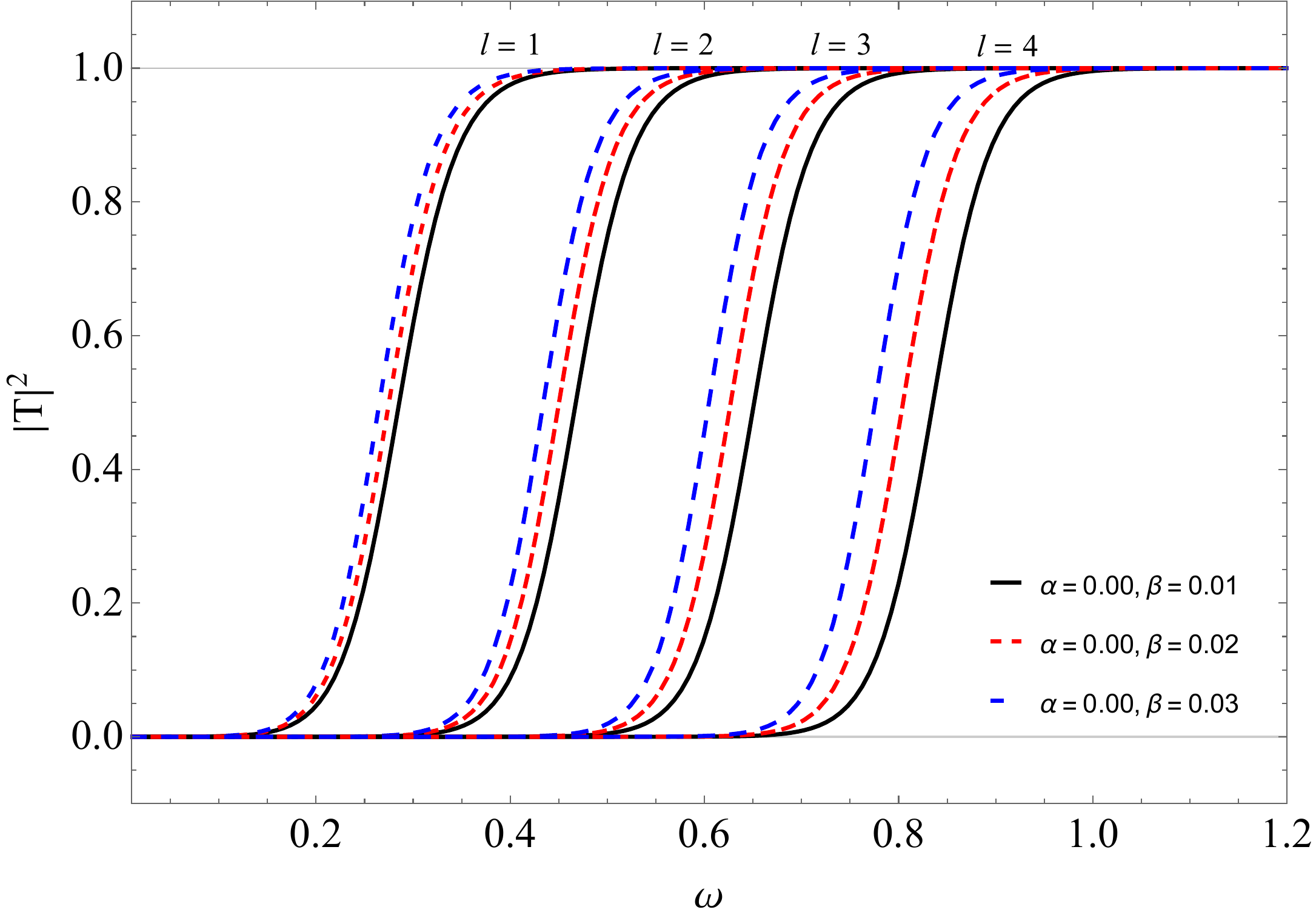}\label{trsa0}}
 \qquad
 \subfigure[]{\includegraphics[scale=0.30]{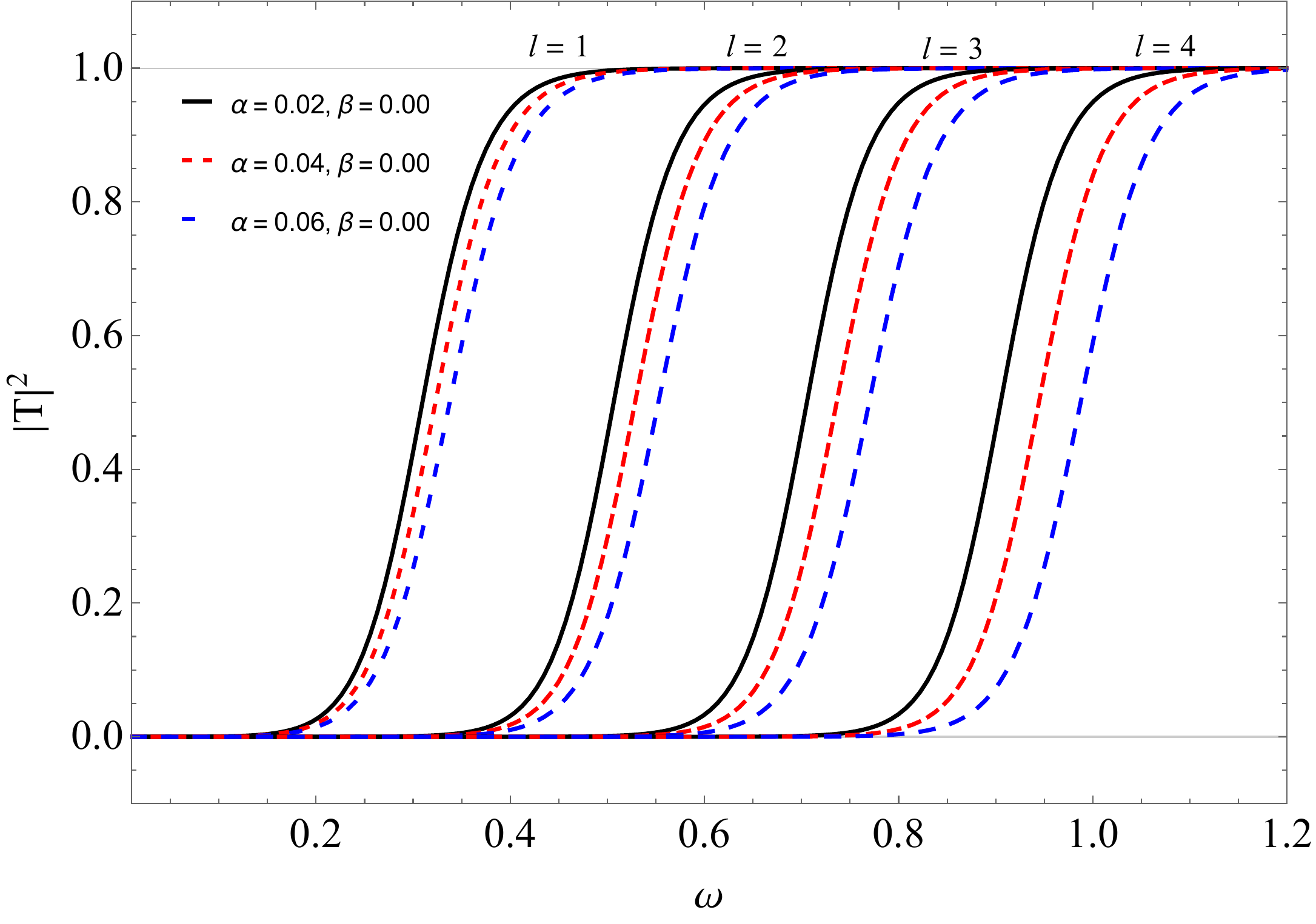}\label{trsb0}}
 \caption{\footnotesize{Transmission coefficient for (a) $ \alpha = 0 $ and $ \beta = 0.01, 0.02 $ and $ 0.03$; while for (b) $ \alpha = 0.02, 0.04 $ and $ 0.06 $ and $ \beta = 0 $ all with $ M = 1 $.}} 
  \label{Tras}
\end{figure}

Now, we will make a brief study on scattering processes by using the WKB method applied to black holes
with GUP~\cite{Konoplya:2019xmn,Konoplya:2019hlu}. 
Thus, we will start by analyzing the boundary conditions applied to the radial equation in terms of the tortoise coordinate. 
Here we have two different situations, the first, when the waves approach the event horizon, and the second, when they move away to infinity.
These two situations are represented in the equations below
\begin{equation}
R_{\omega l} = 
\begin{cases} e^{-i \omega r_{*}} + \mathcal{R} e^{i\omega r_{*}},\qquad r_{*} 
\rightarrow \infty, 
\\
Te^{i\omega r_{*}}, \hspace{2.1cm} r_{*} \rightarrow -\infty,
\end{cases}
\end{equation}
where $ {\cal R} $ and $ T $ represent the reflection and transmission coefficients, respectively. 
At this point, to apply the WKB method to the scattering process, we define $ n+1/2={\cal K} $ 
in equation~\eqref{eqWKB} 
\begin{equation}
\mathcal{K} = \dfrac{i\left(\omega^{2} - V_{0}\right)}{\sqrt{-2V_{0}''}} - \sum_{j=2}^{6}\Omega_{j}(\mathcal{K}).
\label{eqtr}
\end{equation}
Here $ \omega $ is real and the correction terms $\Omega_{j}$ are now functions of $\mathcal{K}$.
Solving equation \eqref{eqtr} for $ {\cal K} $, we can use the result to find the coefficients $ T $ 
and $ {\cal R} $, and so we have the following relations between these terms~\cite{Schutz:1985zz}:
\begin{eqnarray}
|{\cal R}|^{2} & =& \dfrac{1}{1 + e^{-2 i \pi \mathcal{K}}}, \qquad 0 < |{\cal R}|^{2} < 1,
\\ 
|T|^{2} & =& \dfrac{1}{1 + e^{2 i \pi \mathcal{K}}} = 1 - |{\cal R}|^{2}.
\end{eqnarray}
In Fig.~\ref{Tras}, we show the results for the transmission coefficient by applying the WKB approach up to the sixth order. 
Moreover, in order to verify the accuracy of the WKB method, we compared it with the results obtained numerically for the absorption cross section~\cite{Anacleto:2020lel}. 
So in terms of the transmission coefficient $ T $ the absorption cross section is given by
\begin{equation}
\sigma_{abs} = \dfrac{\pi}{\omega^2}\sum_{l = 0}^{\infty}\left(2l + 1\right)|T|^2.
\end{equation}
In Fig.~\ref{abs}, we show the results for the absorption cross section. 
In Fig.~{\ref{absbt}, we have the results considering $ \alpha=0$ and $\beta=0.03 $ 
for the modes $l = 1, 2, 3$ and $4$. 
On the other hand,  in Fig.~\ref{absalp}, we adopted the values of $\alpha = 0.06$ and $ \beta=0 $ for the modes $l = 1, 2$ and $3$. 
So, we have a comparison between the sixth order WKB approach (solid gray line) 
with the numeric (dashed black line) recently obtained in~\cite{Anacleto:2020lel}, and we can see that the lines are exactly overlapping showing the accuracy of the WKB approach with respect to numerical methods for values of $ l > 0 $ 
with different values of $ \alpha $ and $ \beta $.
\begin{figure}[!htb]
 \centering
 \subfigure[]{\includegraphics[scale=0.35]{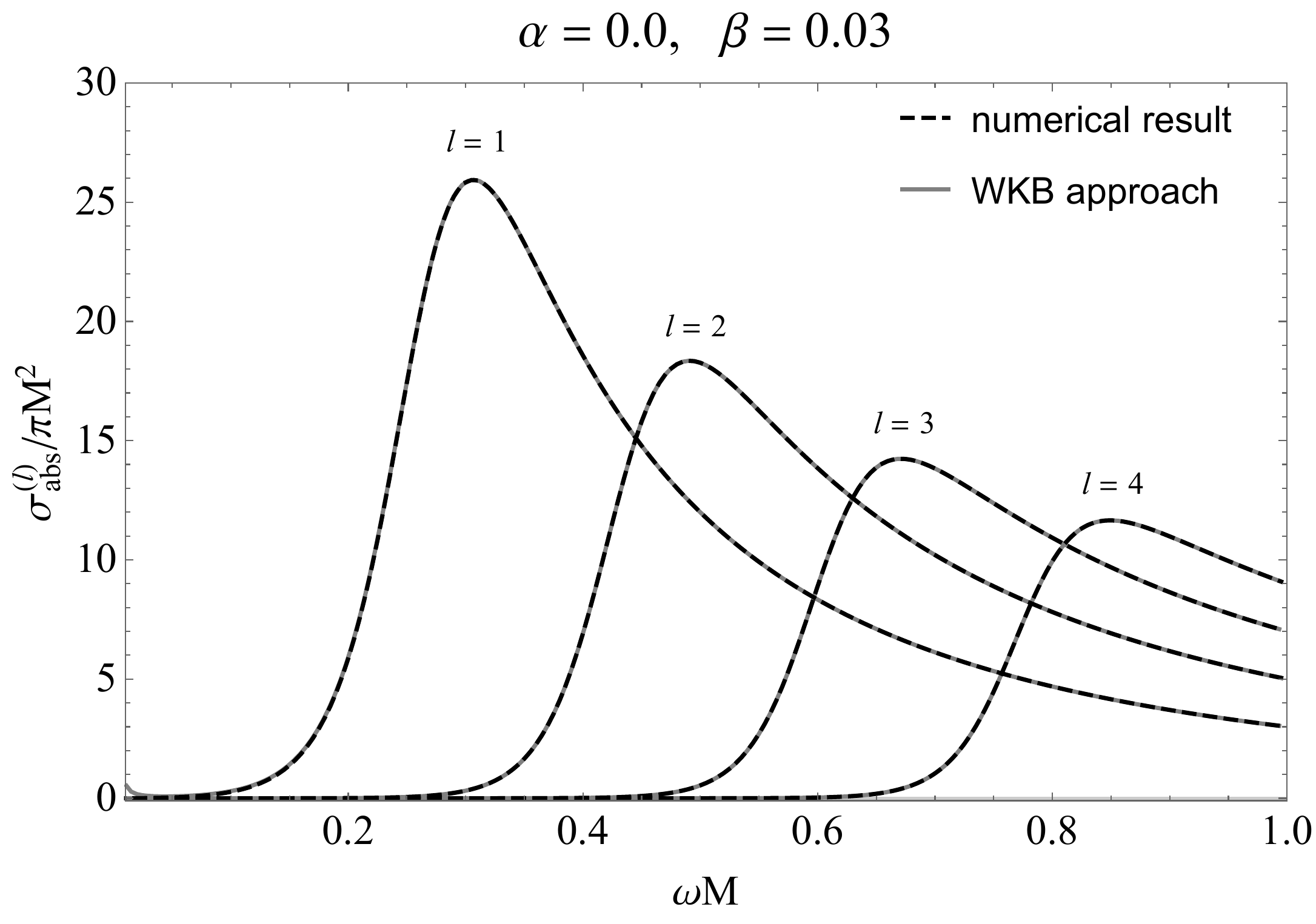}\label{absbt}}
\qquad
 \subfigure[]{\includegraphics[scale=0.35]{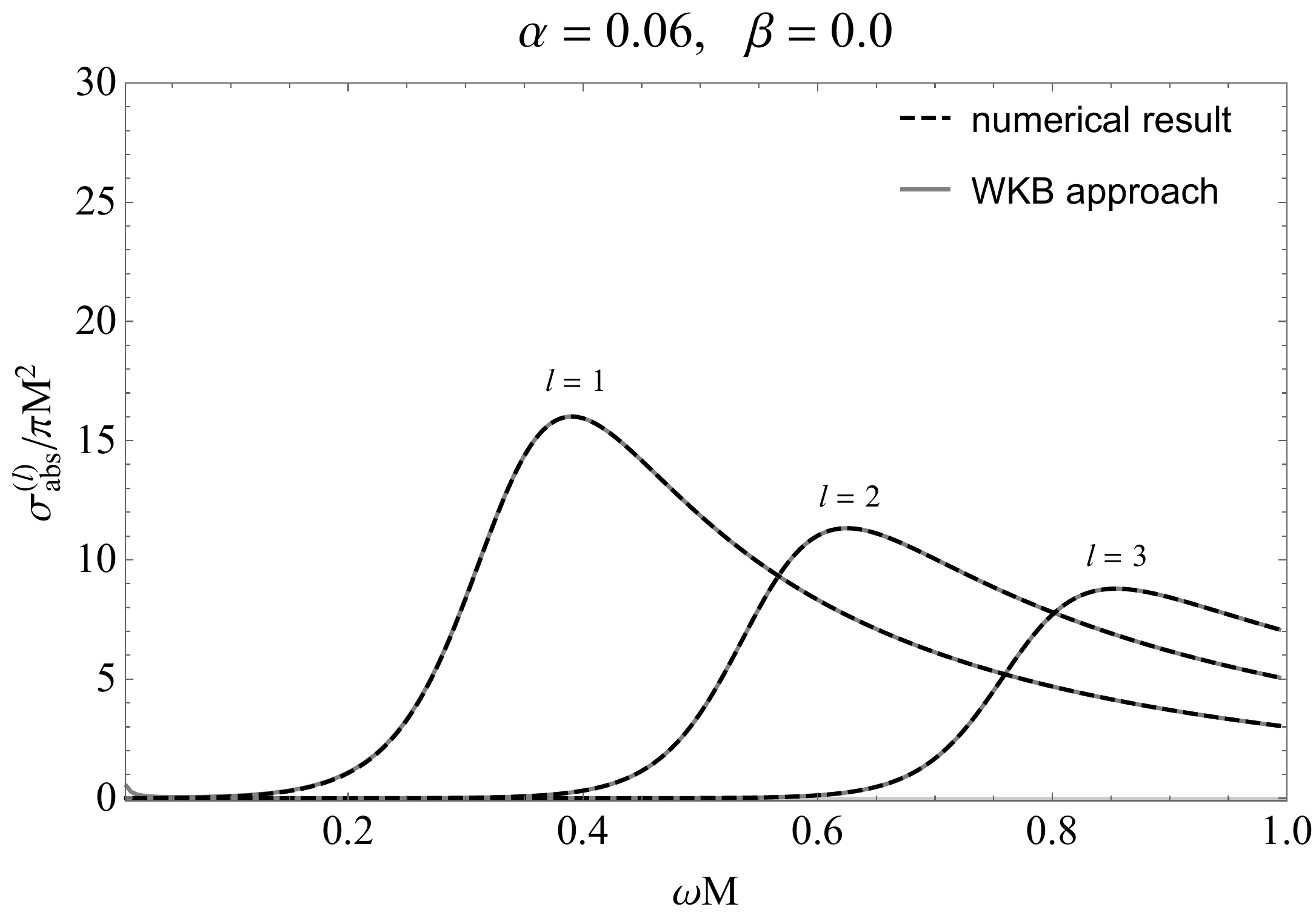}\label{absalp}}
 \caption{\footnotesize{Comparison between the results obtained by the sixth-order WKB and numerical methods for absorption cross section.  }} 
  \label{abs}
\end{figure}

\subsection{Leaver analysis}
We start with rewriting equation (\ref{eqrad}) in terms of the usual radial coordinate in the following form 
\begin{equation}
 r(r - r_{hgup})^{2}\dfrac{d^{2}R}{dr^{2}} + r_{hgup}(r - r_{hgup})\dfrac{dR}{dr} + r^{3}\left(\omega^{2} - V_{eff}\right)R = 0,
\label{ER}
\end{equation}
subject to the following boundary conditions: as $r \rightarrow r_{hgup} $ we have $R\rightarrow (r - r_{hgup})^{-i\omega r_{hgup}}$ and as $r \rightarrow \infty $ we have $R\rightarrow r^{i\omega r_{hgup}}e^{i\omega r}$.
A solution to equation (\ref{ER}) on the horizon ($r = r_{hgup}$) can be written as follows
\begin{equation}
R = \left(\dfrac{r - r_{hgup}}{r}\right)^{-i\omega r_{hgup}}r^{i\omega r_{hgup}}e^{i\omega(r - r_{hgup})}\sum_{k=0}^{\infty}a_{k}\left(\dfrac{r - r_{hgup}}{r}\right)^{k},
\label{anzart}
\end{equation}
and replacing equation (\ref{anzart}) in (\ref{ER}) we find the recurrence relations
\begin{eqnarray}
&& A_{0}a_{1} + B_{0}a_{0} = 0,
\label{cond1}
\\
&& A_{k}a_{k+1} + B_{k}a_{k} + C_{k}a_{k-1} = 0, \qquad k \geq 1 .
\label{cond2}
\end{eqnarray}
The coefficients of the recurrence ratio $A_{k}$, $B_{k}$ and $C_{k}$ are simply functions of $k$, $\omega$, $l$ and the radius $r_{hgup}$. 
\begin{eqnarray}
A_{k} &=& \left[1 + k\right]\left[1 + k - 2ir_{hgup}\omega\right], \\
B_{k}& =& -\left[1 + l(l+1) + 2k+ 2k^{2} - 4ir_{hgup}\omega - 8ikr_{hgup}\omega - 8r_{hgup}^{2}\omega^{2}\right], \\
C_{k} &=& \left[k - 2ir_{hgup}\omega\right]^{2}.
\end{eqnarray}
So in order to determine the $a_k$ coefficients we can write them in terms of a continuous fraction as shown in~\cite{Gautschi}, i.e., 
\begin{equation}
\dfrac{a_{k+1}}{a_{k}}=\dfrac{-C_{k+1}}{B_{k+1} - \dfrac{A_{k+1}C_{k+2}}{B_{k+2}-\dfrac{A_{k+2}C_{k+3}}{B_{k+2} - ...}}},
\end{equation}
which can also be written as follows
\begin{equation}
\dfrac{a_{k+1}}{a_{k}} = \dfrac{-C_{k+1}}{B_{k+1}-}\dfrac{A_{k+1}C_{k+2}}{B_{k+2} -}\dfrac{A_{k+1}C_{k+3}}{B_{k+3} -...}.
\label{efracont}
\end{equation}
At the limit of very large $ k $ we get the following result:
\begin{equation}
\lim\limits_{k \to \infty} \dfrac{a_{k+1}}{a_{k}} = 1 \pm \dfrac{\sqrt{-2i\omega}}{\sqrt{k}} - \dfrac{3 + 4i\omega}{4k}.
\end{equation}
We can obtain the characteristic equation for quasinormal frequencies by making $k=0$
in (\ref{efracont})
and comparing it with the ratio $a_{1}/a_{0}$ obtained in (\ref{cond1})
\begin{equation}
0 = \dfrac{B_{0}}{A_{0}} - \dfrac{C_{1}}{B_{1}-}\dfrac{A_{1}C_{2}}{B_{2} -}\dfrac{A_{1}C_{3}}{B_{3} -...}
\label{efracontcarac}
\end{equation}
This equation defines the quasinormal frequencies $\omega$. To find such frequencies we compute the roots of the equation (\ref{efracontcarac}) via numerical methods. 

\subsection{Results}
At this point, we present the results obtained by the two methods described above and which are shown  
in tables~(\ref{tab1} - \ref{tab4}) below for the modes $ l = 1, 2 $.
In tables~\ref{tab1} and~\ref{tab2}, we see some results for quasinormal modes considering $ M = 1 $. When we set a value for the parameter $ \alpha $ we noticed a decrease in the real and imaginary part of the quasinormal frequency with the increment of $ \beta $. This shows that the quadratic part in these conditions contributes to the decrease of the quasinormal frequency. 
In tables~\ref{tab3} and~\ref{tab4}, we display the results at the small mass limit ($ M = 0.05 $). This result is quite interesting, and also shows the influence of the quadratic part of the GUP ($ \beta \neq0 $).
The quasinormal frequencies are given by $ \omega=\omega_{R} - i\omega_{I} $, where $ \omega_R $ represents the frequency of the oscillation and $ \omega_I $ provides the damping time scale, which implies black hole stability for  $ \omega_I>0$.

Another way to verify these effects of GUP in quasinormal modes and stability is through the graph of frequencies as a function of modes ($ n $) as shown in Fig.~\ref{fig_Qn}, in which we have the real and imagined part of the frequencies for $ l = 1,2,3,4 $ and $ M = 1 $. 
We have the graphics from left to right an addition of the parameter $ \alpha$. 
This causes a slight change in the curves that is more perceptible for $ l = 1 $, while the addition of the parameter $ \beta $ causes a change being more noticeable for the imaginary part of the frequency mainly for large $ n $. 
Note that by increasing $l$ and the parameter $\beta$ the real and imaginary part approach the horizontal axis but they never cross the axis, i.e., there are no changes  of sign. This effect is most noticeable for the imaginary part of the frequency.  
Thus, the result indicates that the black hole keeps stable under the scalar field perturbation.

In Fig.~\ref{plancomplex} we have the results of the quasinormal frequencies in a complex plane for three multipole families $l = 0, 1, 2$. In Fig.~\ref{plancomplex1} we consider only the quadratic part ($\alpha = 0$). By comparing with the Schwarzschild case (black curves), the curves move to the left with the increase of the parameter $\beta$, while for the linear part only ($\beta=0$), as shown in Fig.~\ref{plancomplex2}, the curves move to the right with respect to the Schwarzschild case when we increase the parameter $\alpha$.
\\
\begin{table}[h!]
\begin{footnotesize}
\begin{center}
\caption{\footnotesize{ Quasinormal frequencies for the $l=1$ mode, considering the linear and quadratic GUP.}} 
\label{tab1}
\begin{tabular}{c||c||c|c|c|c|c|c}
\hline
\multicolumn{2} {c|}{$M=1$}&\multicolumn{2} {|c|}{ $\omega_{0}$ } & \multicolumn{2} {|c|}{ $\omega_{1}$ } & \multicolumn{2} {|c}{ $\omega_{2}$ } \\
\hline
 $\alpha $ & $\beta$   &  6th order WKB &  numerical & 6th order WKB & numerical & 6th order WKB & numerical\\
 \hline
\multirow{4}{*}{$ 0.00 $} 
   & 0.00  & 0.292910 - 0.097762i & 0.292936 - 0.097660i & 0.264471 - 0.306518i & 0.264449 - 0.306257i & 0.231014 - 0.542166i & 0.229539 - 0.540133i \\
   & 0.01  & 0.281644 - 0.094002i & 0.281669 - 0.093904i & 0.254299 - 0.294729i & 0.254278 - 0.294478i & 0.222129 - 0.521313i & 0.220711 - 0.519359i \\
   & 0.02  & 0.271213 - 0.090520i & 0.271237 - 0.090426i & 0.244881 - 0.283813i & 0.244860 - 0.283572i & 0.213902 - 0.502005i & 0.212536 - 0.500124i \\
   & 0.03  & 0.261526 - 0.087287i & 0.261550 - 0.087196i & 0.236135 - 0.273677i & 0.236115 - 0.273444i & 0.206263 - 0.484076i & 0.204946 - 0.482262i \\
   \hline
   \hline
\multirow{4}{*}{$ 0.02 $}
   & 0.00  & 0.305114 - 0.101835i & 0.305142 - 0.101729i & 0.275491 - 0.319290i & 0.275467 - 0.319018i & 0.240640 - 0.564755i & 0.239103 - 0.562639i  \\
   & 0.01  & 0.292910 - 0.097762i & 0.292936 - 0.097660i & 0.264471 - 0.306518i & 0.264449 - 0.306257i & 0.231014 - 0.542165i & 0.229539 - 0.540133i \\
   & 0.02  & 0.281644 - 0.094002i & 0.281669 - 0.093904i & 0.254299 - 0.294729i & 0.254278 - 0.294478i & 0.222128 - 0.521315i & 0.220711 - 0.519359i \\
   & 0.03  & 0.271213 - 0.090520i & 0.271237 - 0.090426i & 0.244881 - 0.283813i & 0.244860 - 0.283572i & 0.213902 - 0.502005i & 0.212536 - 0.500124i \\
   \hline
   \hline
\multirow{4}{*}{$ 0.04 $}
   & 0.00  & 0.318380 - 0.106263i & 0.318409 - 0.106152i & 0.287469 - 0.333172i & 0.287444 - 0.332888i & 0.251103 - 0.589309i & 0.249499 - 0.587102i \\
   & 0.01  & 0.305114 - 0.101835i & 0.305142 - 0.101729i & 0.275491 - 0.319290i & 0.275467 - 0.319018i & 0.240640 - 0.564755i & 0.239103 - 0.562639i \\
   & 0.02  & 0.292910 - 0.097762i & 0.292936 - 0.097660i & 0.264471 - 0.306518i & 0.264449 - 0.306257i & 0.231014 - 0.542165i & 0.229539 - 0.540133i \\
   & 0.03  & 0.281644 - 0.094002i & 0.281669 - 0.093904i & 0.254299 - 0.294730i & 0.254278 - 0.294478i & 0.222128 - 0.521315i & 0.220711 - 0.519359i \\
   \hline
   \hline
\multirow{4}{*}{$ 0.06 $}
   & 0.00  & 0.332852 - 0.111093i & 0.332882 - 0.110977i & 0.300536 - 0.348315i & 0.300510 - 0.348020i & 0.262518 - 0.616093i & 0.260840 - 0.613788i \\
   & 0.01  & 0.318380 - 0.106263i & 0.318409 - 0.106152i & 0.287469 - 0.333171i & 0.287444 - 0.332888i & 0.251104 - 0.589306i & 0.249499 - 0.587102i \\
   & 0.02  & 0.305114 - 0.101835i & 0.305142 - 0.101729i & 0.275491 - 0.319290i & 0.275467 - 0.319018i & 0.240640 - 0.564755i & 0.239103 - 0.562639i \\
   & 0.03  & 0.292910 - 0.097762i & 0.292936 - 0.097660i & 0.264471 - 0.306518i & 0.264449 - 0.306257i & 0.231014 - 0.542165i & 0.229539 - 0.540133i \\
 \hline
 \end{tabular}
\end{center}
\end{footnotesize}
\end{table}

\begin{table}[h!]
\begin{footnotesize}
\begin{center}
\caption{\footnotesize{ Quasinormal frequencies for the $l = 2$ mode, considering the linear and quadratic GUP.}} 
\label{tab2}
\begin{tabular}{c||c||c|c|c|c|c|c}
\hline
\multicolumn{2} {c|}{$M = 1$}&\multicolumn{2} {|c|}{ $\omega_{0}$ } & \multicolumn{2} {|c|}{ $\omega_{1}$ } & \multicolumn{2} {|c}{ $\omega_{2}$ } \\
\hline
 $\alpha $ & $\beta$   &  6th order WKB &  numerical & 6th order WKB & numerical & 6th order WKB & numerical\\
 \hline
 \multirow{4}{*}{$ 0.00 $} 
   & 0.00  & 0.483642 - 0.096766i & 0.483644 - 0.0967588i & 0.463847 - 0.295627i & 0.463851 - 0.295604i & 0.430386 - 0.508700i & 0.430544 - 0.508558i \\
   & 0.01  & 0.465040 - 0.093044i & 0.465042 - 0.0930373i & 0.446006 - 0.284257i & 0.446010 - 0.284235i & 0.413833 - 0.489134i & 0.413985 - 0.488998i \\
   & 0.02  & 0.447817 - 0.089598i & 0.447818 - 0.0895915i & 0.429488 - 0.273729i & 0.429491 - 0.273707i & 0.398505 - 0.471018i & 0.398652 - 0.470887i \\
   & 0.03  & 0.431823 - 0.086398i & 0.431825 - 0.0863918i & 0.414149 - 0.263953i & 0.414152 - 0.263932i & 0.384273 - 0.454196i & 0.384414 - 0.454070i \\
   \hline
   \hline
 \multirow{4}{*}{$ 0.02 $}
   & 0.00  & 0.503794 - 0.100798i & 0.503796 - 0.100790i & 0.483174 - 0.307945i & 0.483178 - 0.307921i & 0.448318 - 0.529896i & 0.448483 - 0.529748i \\
   & 0.01  & 0.483642 - 0.096766i & 0.483644 - 0.096759i & 0.463847 - 0.295627i & 0.463851 - 0.295604i & 0.430386 - 0.508700i & 0.430544 - 0.508558i \\
   & 0.02  & 0.465040 - 0.093044i & 0.465042 - 0.093037i & 0.446006 - 0.284257i & 0.446010 - 0.284235i & 0.413832 - 0.489135i & 0.413985 - 0.488998i \\
   & 0.03  & 0.447817 - 0.089598i & 0.447818 - 0.089591i & 0.429488 - 0.273729i & 0.429491 - 0.273707i & 0.398505 - 0.471018i & 0.398652 - 0.470887i \\
   \hline
   \hline
  \multirow{4}{*}{$ 0.04 $}
   & 0.00  & 0.525698 - 0.105181i & 0.525700 - 0.105173i & 0.504181 - 0.321334i & 0.504185 - 0.321309i & 0.467811 - 0.552935i & 0.467983 - 0.552781i \\
   & 0.01  & 0.503794 - 0.100798i & 0.503796 - 0.100790i & 0.483174 - 0.307945i & 0.483178 - 0.307921i & 0.448318 - 0.529896i & 0.448483 - 0.529748i \\
   & 0.02  & 0.483642 - 0.096766i & 0.483644 - 0.096759i & 0.463847 - 0.295627i & 0.463851 - 0.295604i & 0.430386 - 0.508700i & 0.430544 - 0.508558i \\
   & 0.03  & 0.465040 - 0.093044i & 0.465042 - 0.093037i & 0.446006 - 0.284257i & 0.446010 - 0.284235i & 0.413832 - 0.489135i & 0.413985 - 0.488998i \\
   \hline
   \hline
  \multirow{4}{*}{$ 0.06 $}
   & 0.00  & 0.549593 - 0.109961i & 0.549595 - 0.109953i & 0.527099 - 0.335940i & 0.527103 - 0.335914i & 0.489075 - 0.578068i & 0.489255 - 0.577907i \\
   & 0.01  & 0.525698 - 0.105181i & 0.525700 - 0.105173i & 0.504181 - 0.321334i & 0.504185 - 0.321309i & 0.467811 - 0.552935i & 0.467983 - 0.552781i \\
   & 0.02  & 0.503794 - 0.100798i & 0.503796 - 0.100790i & 0.483174 - 0.307945i & 0.483178 - 0.307921i & 0.448319 - 0.529895i & 0.448483 - 0.529748i \\
   & 0.03  & 0.483642 - 0.096766i & 0.483644 - 0.096759i & 0.463847 - 0.295627i & 0.463851 - 0.295604i & 0.430386 - 0.508700i & 0.430544 - 0.508558i \\
 \hline
 \end{tabular}
\end{center}
\end{footnotesize}
\end{table}

\begin{table}[h!]
\begin{footnotesize}
\begin{center}
\caption{\footnotesize{ Quasinormal frequencies at the small mass limit for $l = 1$ mode. }} 
\label{tab3}
\begin{tabular}{c||c||c|c|c|c|c|c}
\hline
\multicolumn{2} {c|}{$M = 0.05$}&\multicolumn{2} {|c|}{ $\omega_{0}$ } & \multicolumn{2} {|c|}{ $\omega_{1}$ } & \multicolumn{2} {|c}{ $\omega_{2}$ } \\
\hline
 $\alpha $ & $\beta$   &  6th order WKB &  numerical & 6th order WKB & numerical & 6th order WKB & numerical\\
 \hline
 \multirow{3}{*}{$ 0.00 $} 
   & 0.01  & 0.344600 - 0.115014i & 0.344631 - 0.114894i & 0.311142 - 0.360610i & 0.311116 - 0.360303i & 0.271782 - 0.637841i & 0.270046 - 0.635451i \\
   & 0.02  & 0.177537 - 0.059188i & 0.177537 - 0.059188i & 0.160286 - 0.185769i & 0.160272 - 0.185611i & 0.134603 - 0.481920i & 0.123187 - 0.477756i \\
   & 0.03  & 0.119555 - 0.039903i & 0.119566 - 0.039861i & 0.107947 - 0.125109i & 0.107938 - 0.125003i & 0.094292 - 0.221292i & 0.093689 - 0.220463i \\
   \hline
   \hline
 \multirow{3}{*}{$ 0.03 $}
   & 0.01  & 0.370772 - 0.123749i & 0.370805 - 0.123620i & 0.334774 - 0.387998i & 0.334745 - 0.387668i & 0.292423 - 0.686285i & 0.290556 - 0.683713i \\
   & 0.02  & 0.184220 - 0.061485i & 0.184237 - 0.061421i & 0.166334 - 0.192779i & 0.166320 - 0.192615i & 0.145292 - 0.340984i & 0.144364 - 0.339707i \\
   & 0.03  & 0.122556 - 0.040904i & 0.122567 - 0.040862i & 0.110657 - 0.128250i & 0.110648 - 0.128141i & 0.096659 - 0.226847i & 0.096042 - 0.225997i \\
   \hline
   \hline
 \multirow{3}{*}{$ 0.06 $}
   & 0.01  & 0.401246 - 0.133920i & 0.401282 - 0.133781i & 0.362291 - 0.419885i & 0.362258 - 0.419531i & 0.316464 - 0.742678i & 0.314437 - 0.739909i  \\
   & 0.02  & 0.191444 - 0.063897i & 0.191462 - 0.063830i & 0.172857 - 0.200339i & 0.172842 - 0.200168i & 0.150990 - 0.354357i & 0.150026 - 0.353028i \\
   & 0.03  & 0.125712 - 0.041958i & 0.125724 - 0.041914i & 0.113507 - 0.131553i & 0.113497 - 0.131441i & 0.099148 - 0.232689i & 0.0985147 - 0.231817i \\
 \hline
 \end{tabular}
\end{center}
\end{footnotesize}
\end{table}

\begin{table}[h!]
\begin{footnotesize}
\begin{center}
\caption{\footnotesize{ Quasinormal frequencies at the small mass limit for $l = 2$ mode.}} 
\label{tab4}
\begin{tabular}{c||c||c|c|c|c|c|c}
\hline
\multicolumn{2} {c|}{$M = 0.05$}&\multicolumn{2} {|c|}{ $\omega_{0}$ } & \multicolumn{2} {|c|}{ $\omega_{1}$ } & \multicolumn{2} {|c}{ $\omega_{2}$ } \\
\hline
 $\alpha $ & $\beta$   &  6th order WKB &  numerical & 6th order WKB & numerical & 6th order WKB & numerical\\
 \hline
 \multirow{3}{*}{$ 0.00 $} 
   & 0.01  & 0.568990 - 0.113842i & 0.568993 - 0.113834i & 0.545702 - 0.347796i & 0.545707 - 0.347769i & 0.506336 - 0.598470i & 0.506522 - 0.598304i \\
   & 0.02  & 0.293116 - 0.058646i & 0.293117 - 0.058642i & 0.281119 - 0.179168i & 0.281122 - 0.179154i & 0.260840 - 0.308303i & 0.260936 - 0.308217i \\
   & 0.03  & 0.197405 - 0.039496i & 0.197406 - 0.0394934i & 0.189325 - 0.120664i & 0.189327 - 0.120655  & 0.175668 - 0.207633i & 0.175732 - 0.207575i \\
   \hline
   \hline
 \multirow{3}{*}{$ 0.03 $}
   & 0.01  & 0.612205 - 0.122489i & 0.612207 - 0.122479i & 0.587148 - 0.374211i & 0.587153 - 0.374182i & 0.544792 - 0.643924i & 0.544992 - 0.643745i \\
   & 0.02  & 0.304177 - 0.060859i & 0.304179 - 0.060855i & 0.291728 - 0.185929i & 0.291730 - 0.185914i & 0.270683 - 0.319937i & 0.270782 - 0.319848i \\
   & 0.03  & 0.202361 - 0.040488i & 0.202361 - 0.0404848i & 0.194078 - 0.123693i & 0.19408 - 0.123684i & 0.180078 - 0.212845i & 0.180144 - 0.212786 \\
   \hline
   \hline
 \multirow{3}{*}{$ 0.06 $}
   & 0.01  & 0.662523 - 0.132556i & 0.662526 - 0.132546i & 0.635407 - 0.404968i & 0.635412 - 0.404937i & 0.589570 - 0.696849i & 0.589786 - 0.696655i \\
   & 0.02  & 0.316106 - 0.063246i & 0.316107 - 0.063241i & 0.303168 - 0.193220i & 0.303170 - 0.193205i & 0.281298 - 0.332484i & 0.281401 - 0.332391i \\
   & 0.03  & 0.207572 - 0.041530i & 0.207572 - 0.041527i & 0.199076 - 0.126879i & 0.199078 - 0.126869i & 0.184715 - 0.218326i & 0.184783 - 0.218265i \\
 \hline
 \end{tabular}
\end{center}
\end{footnotesize}
\end{table}

\begin{figure}[!htb]
 \centering
 \subfigure[]{\includegraphics[scale=0.30]{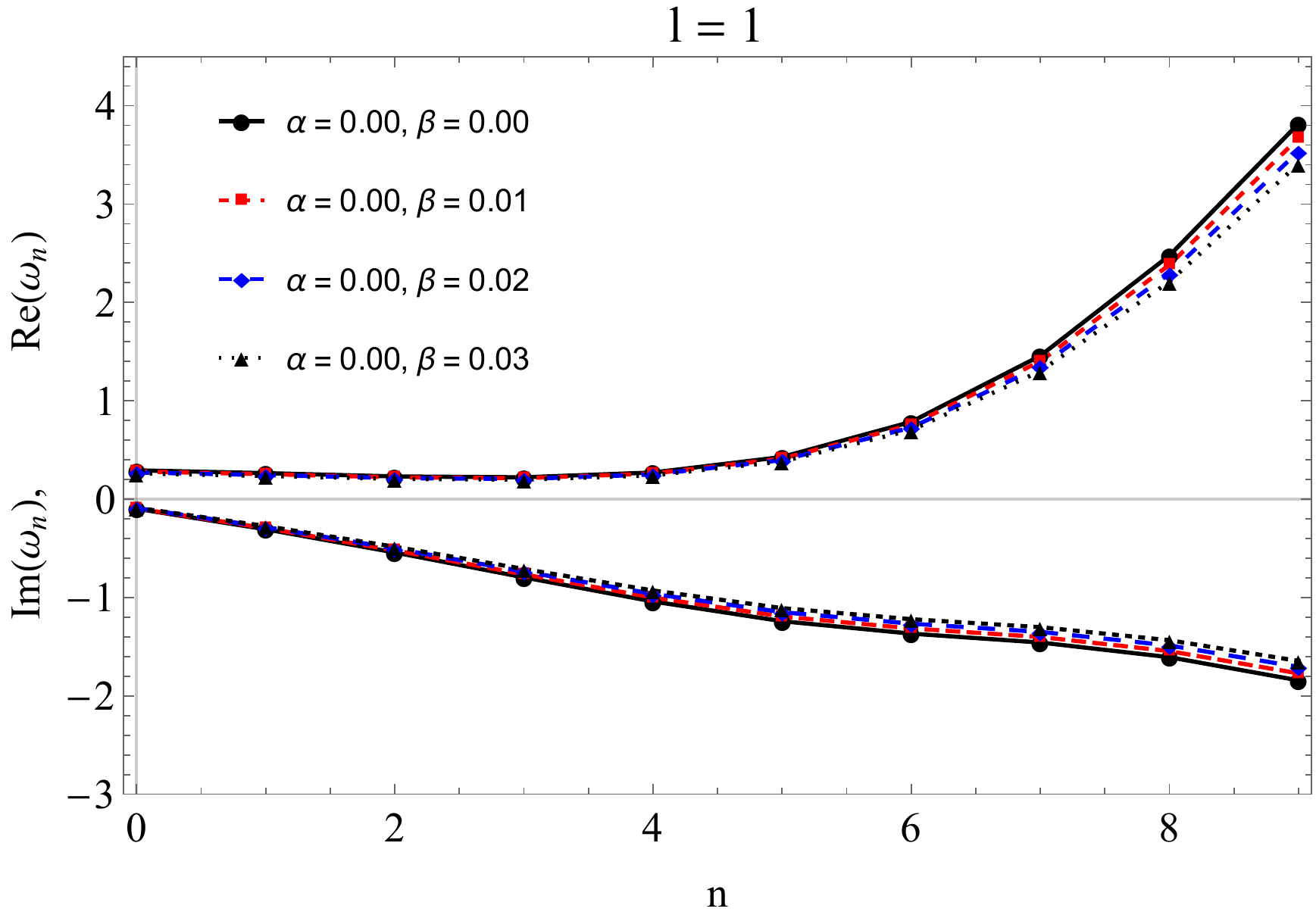}\label{figQN1l1}}
 \qquad
 \subfigure[]{\includegraphics[scale=0.30]{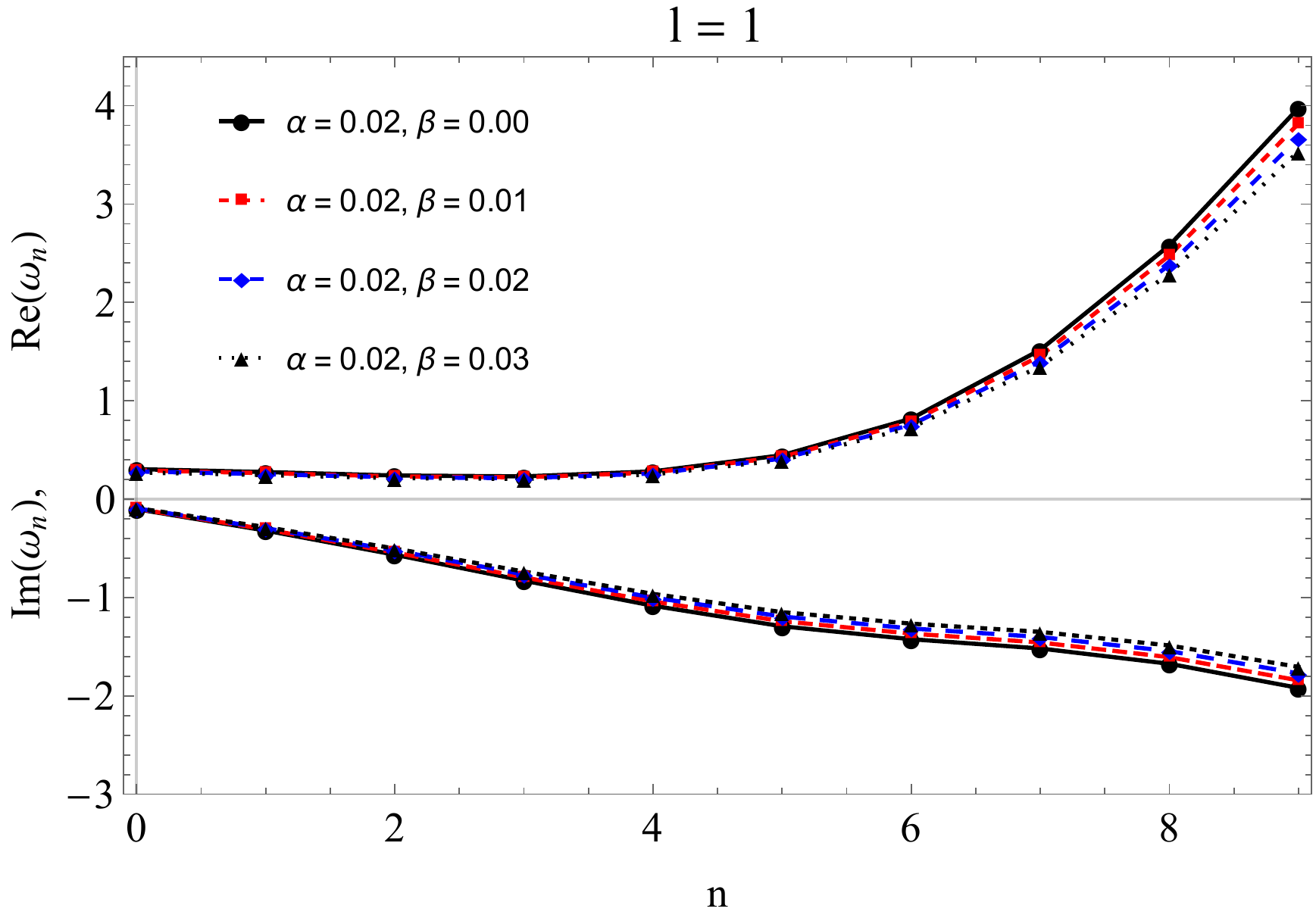}\label{figQN2l1}}
 \qquad
 \subfigure[]{\includegraphics[scale=0.30]{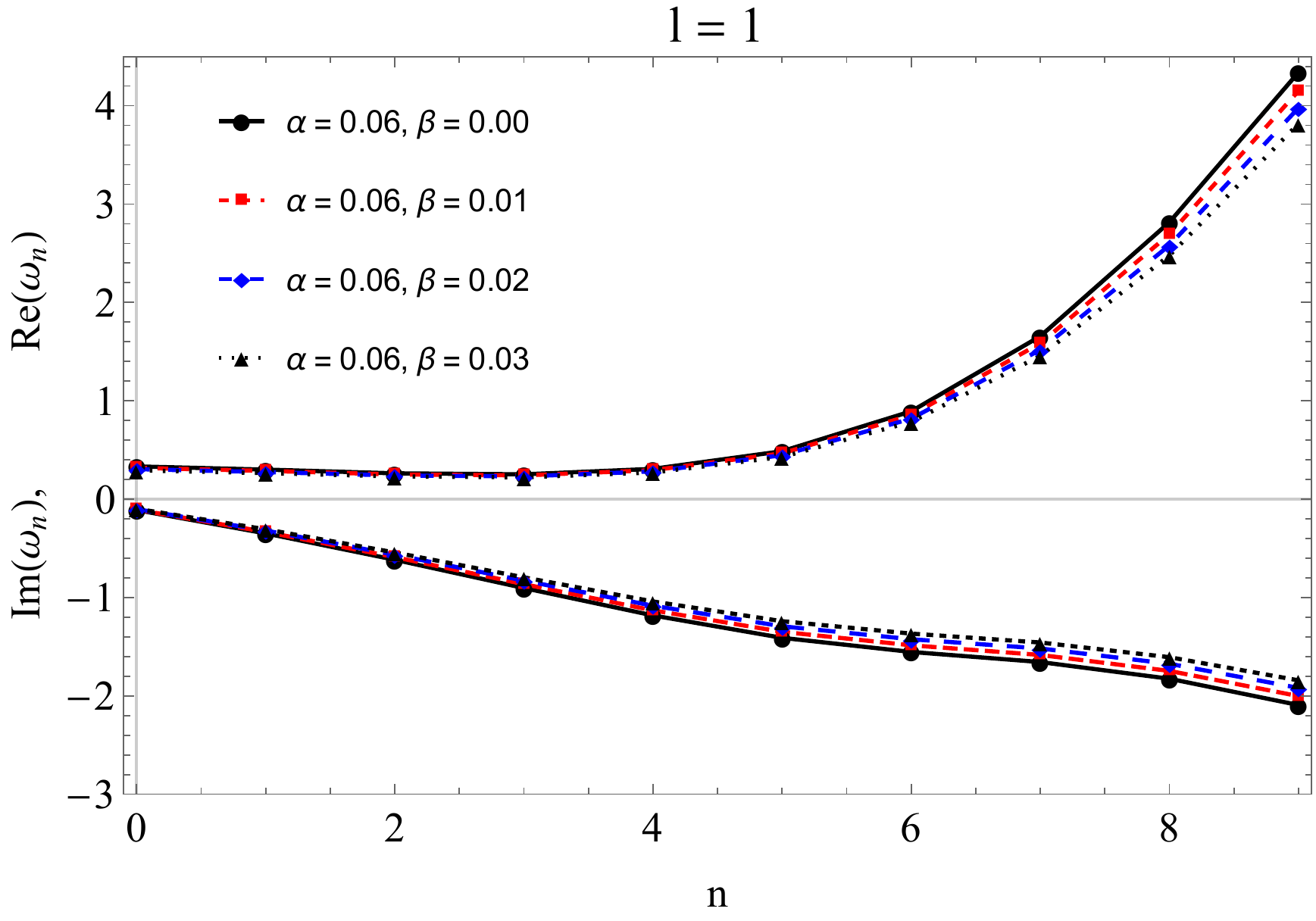}\label{figQN3l1}}
 \qquad
 \subfigure[]{\includegraphics[scale=0.30]{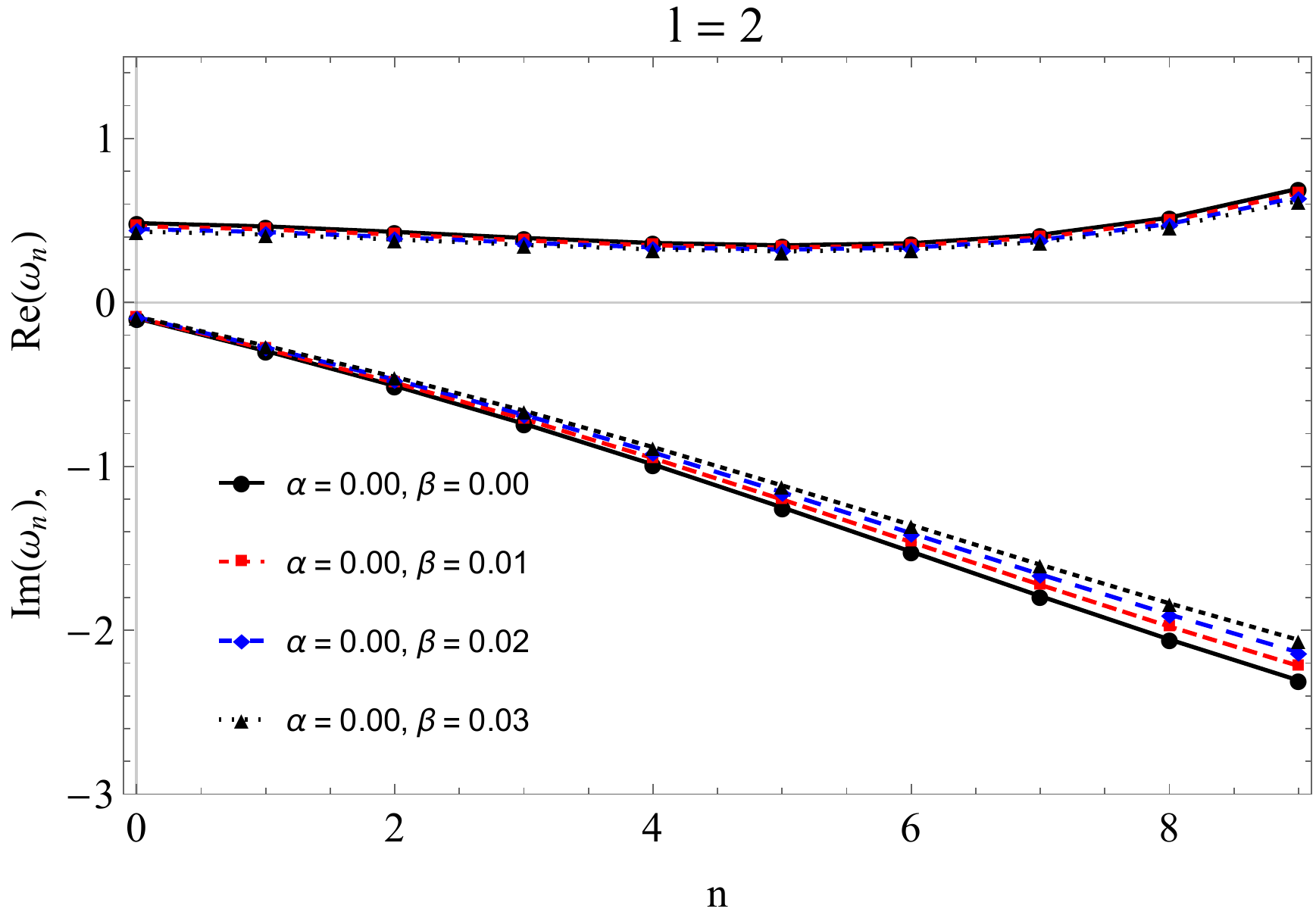}\label{figQN1l2}}
 \qquad
 \subfigure[]{\includegraphics[scale=0.30]{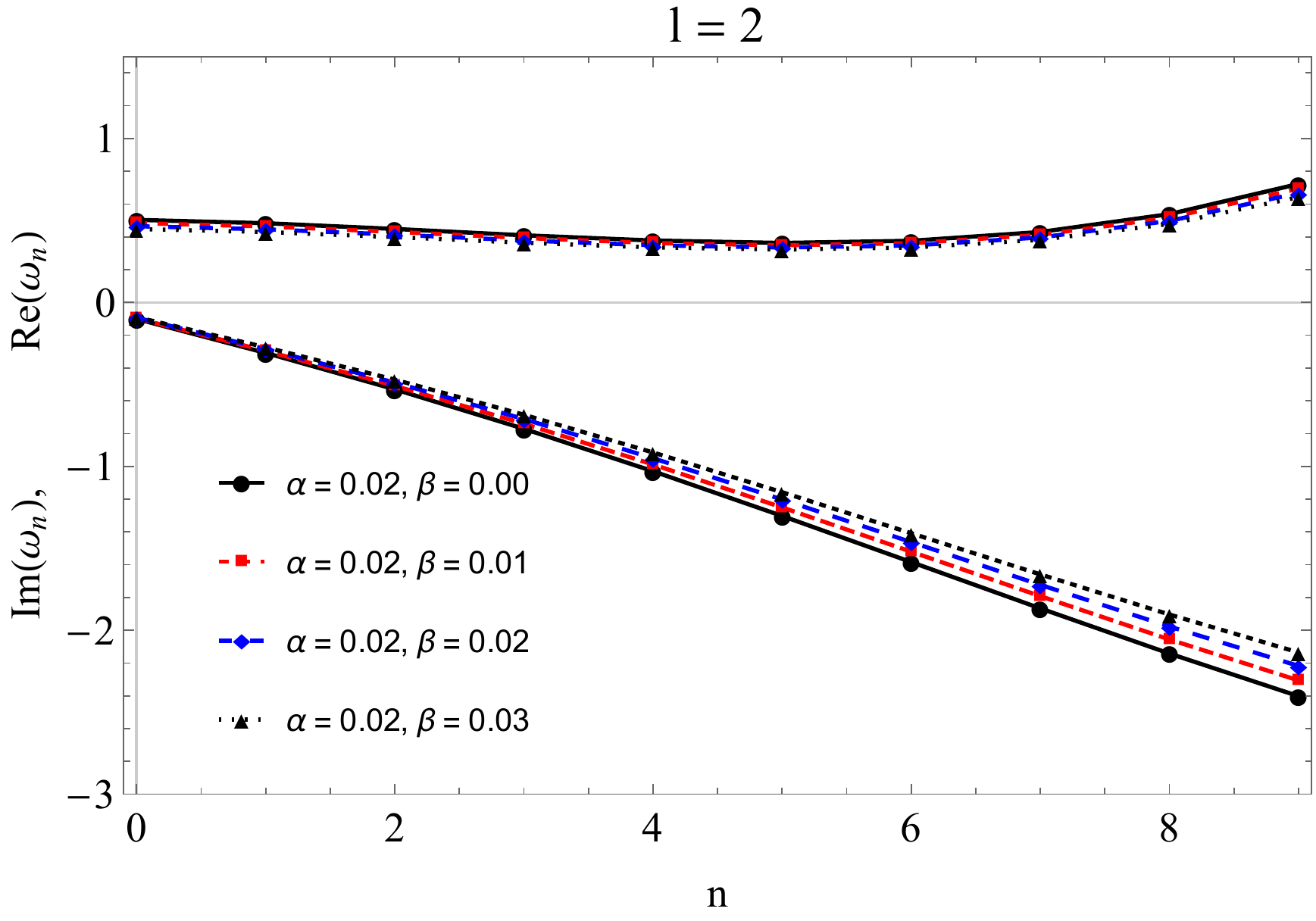}\label{figQN2l2}}
 \qquad
 \subfigure[]{\includegraphics[scale=0.30]{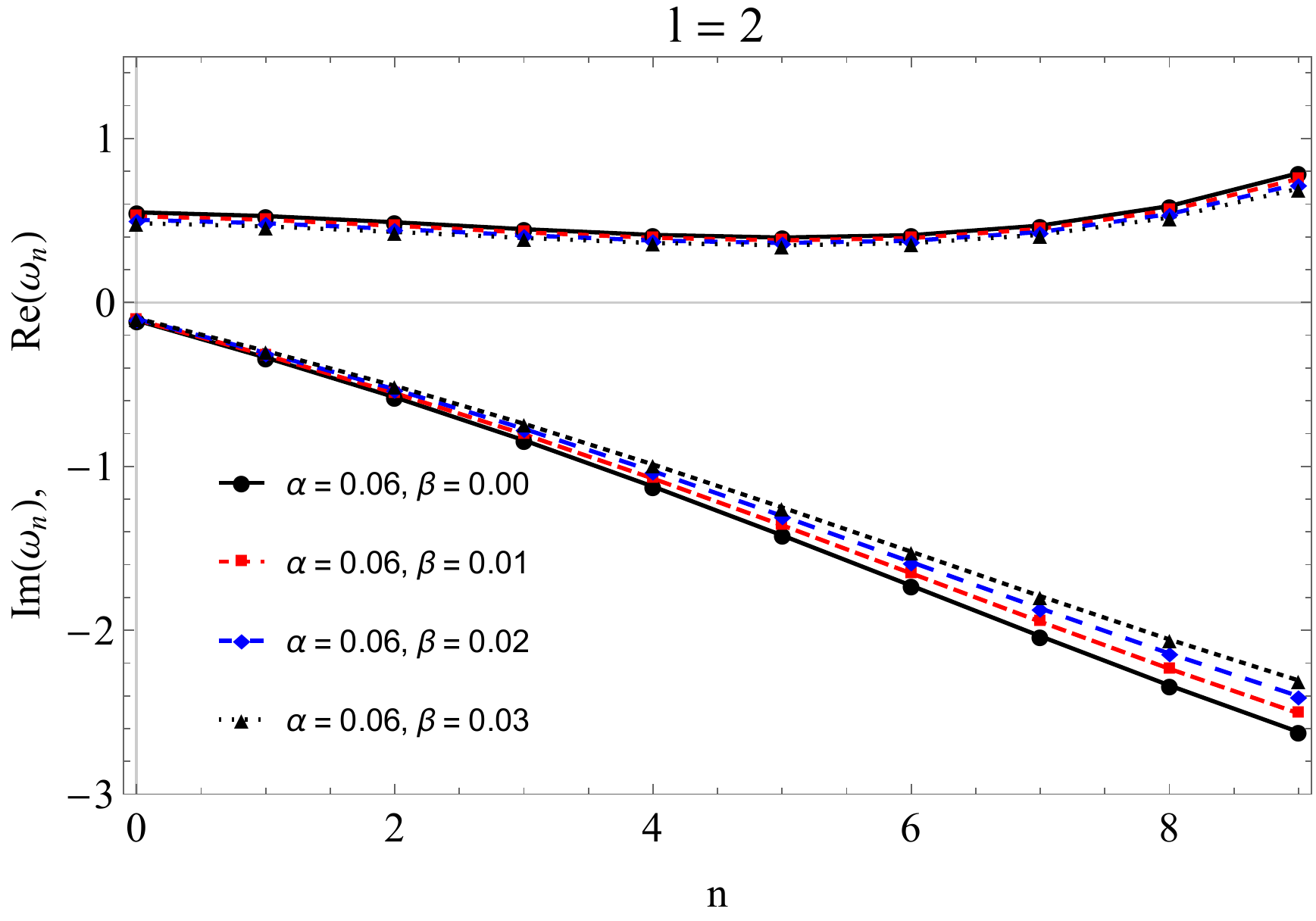}\label{figQN3l2}}
 \qquad
 \subfigure[]{\includegraphics[scale=0.30]{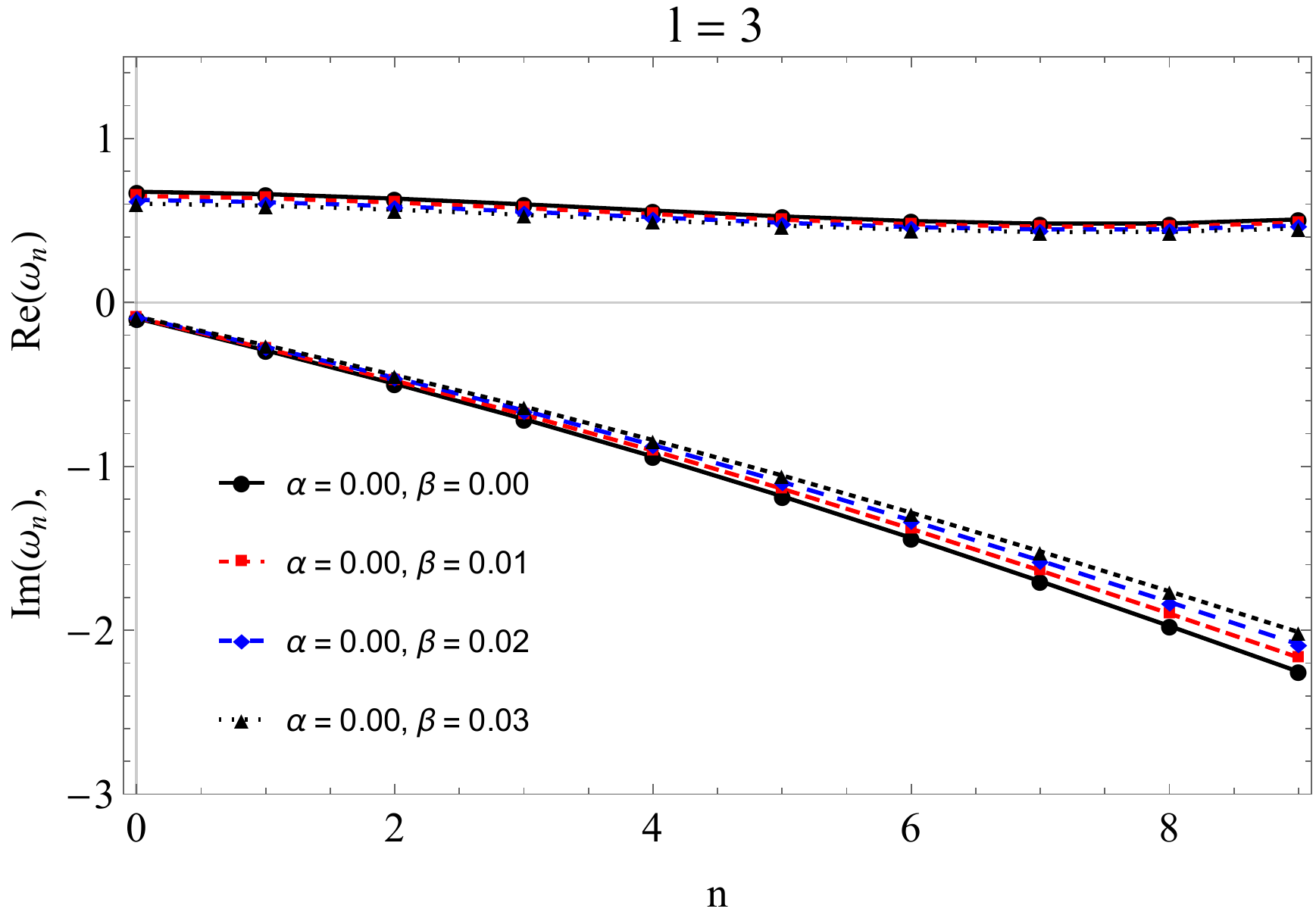}\label{figQN1l3}}
 \qquad
 \subfigure[]{\includegraphics[scale=0.30]{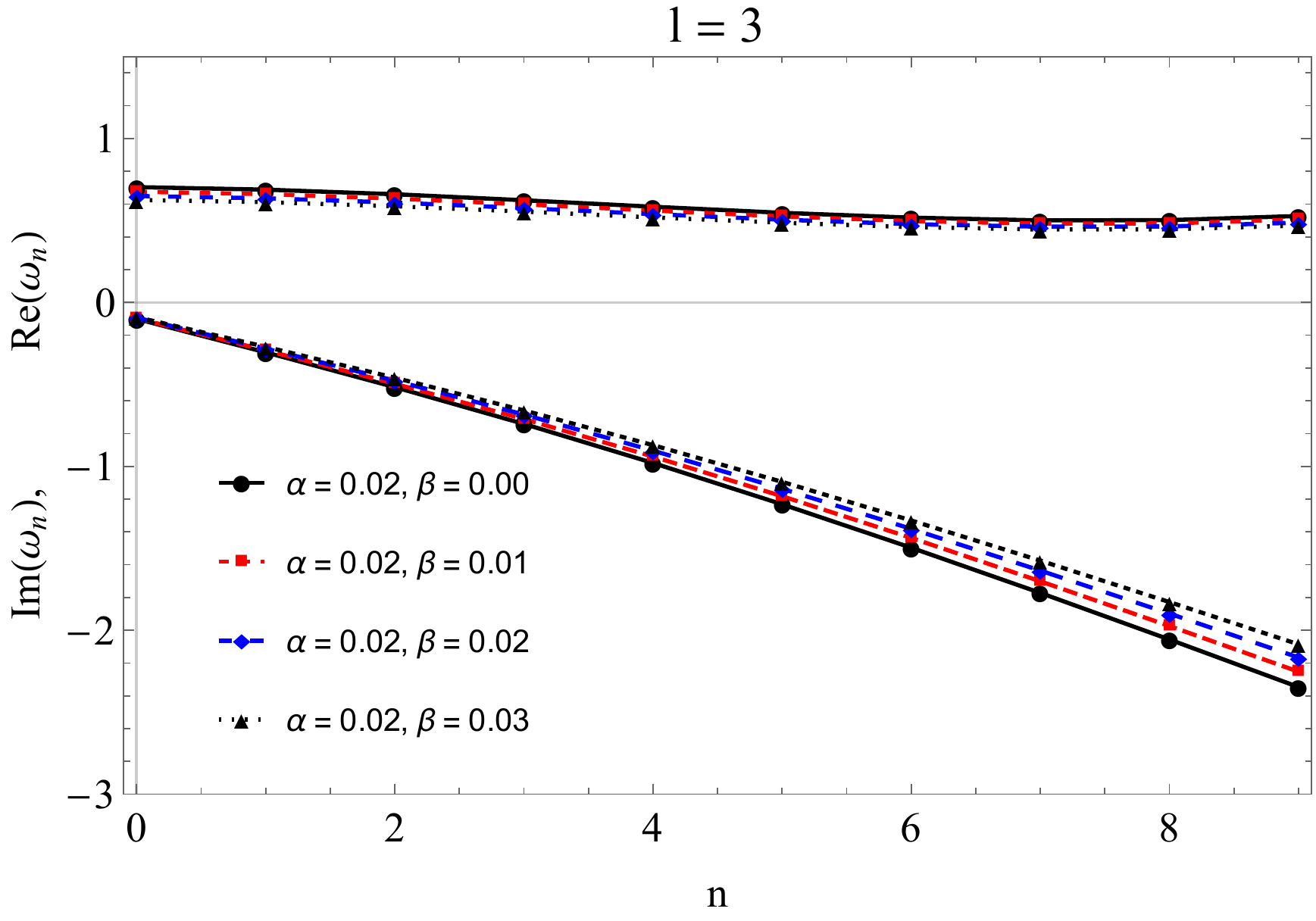}\label{figQN2l3}}
 \qquad
 \subfigure[]{\includegraphics[scale=0.30]{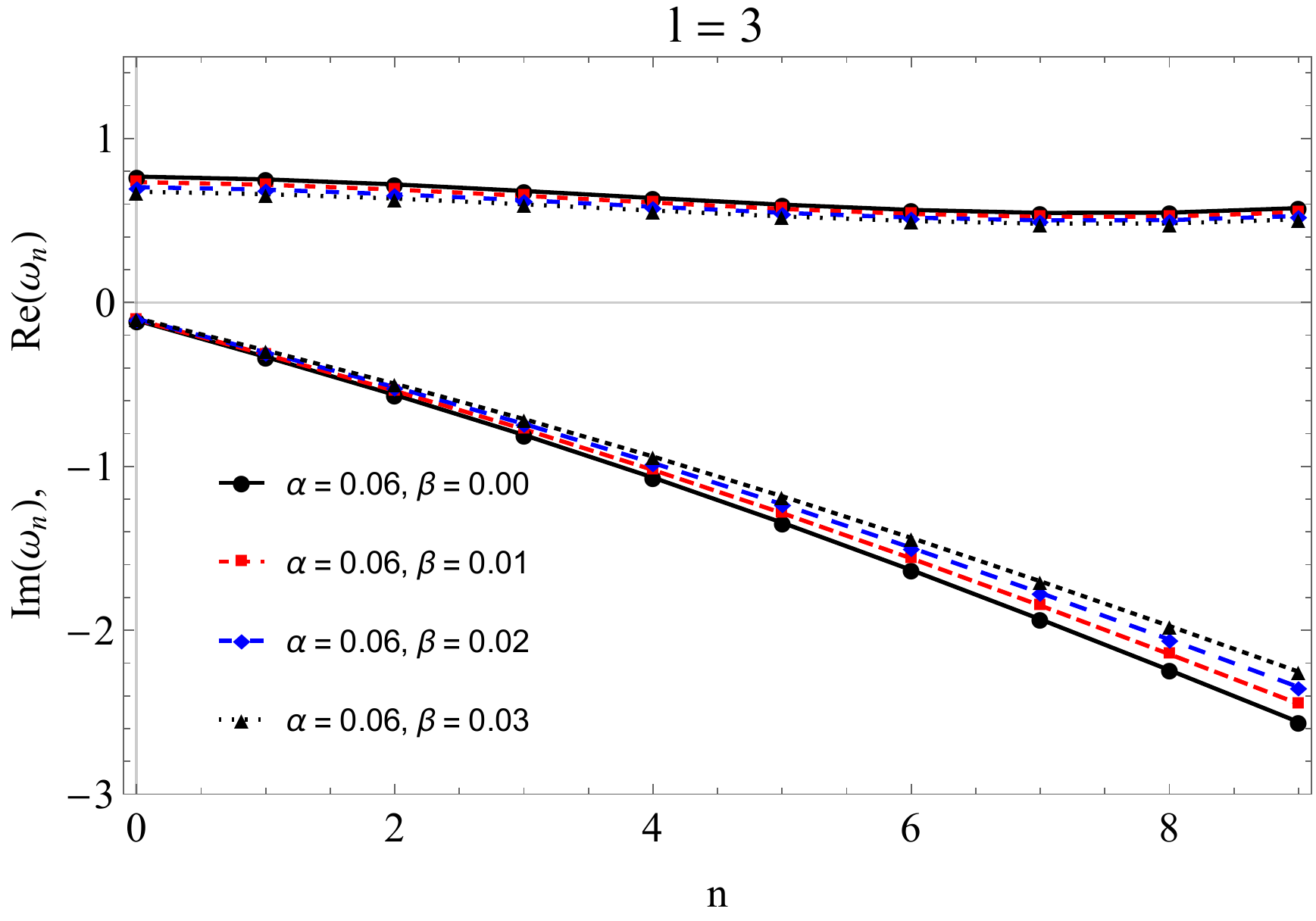}\label{figQN3l3}}
 \qquad
 \subfigure[]{\includegraphics[scale=0.30]{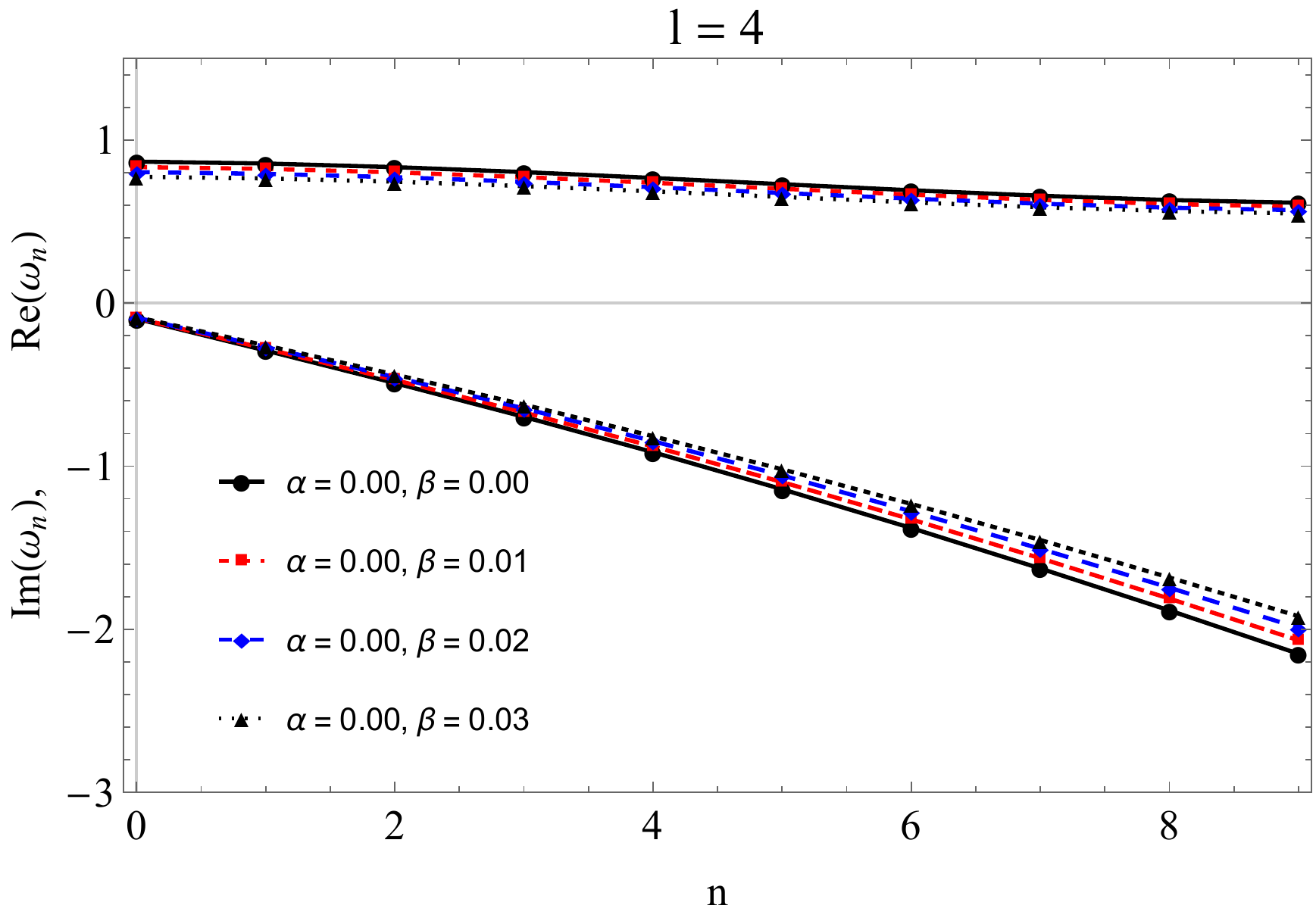}\label{figQN1l4}}
 \qquad
 \subfigure[]{\includegraphics[scale=0.30]{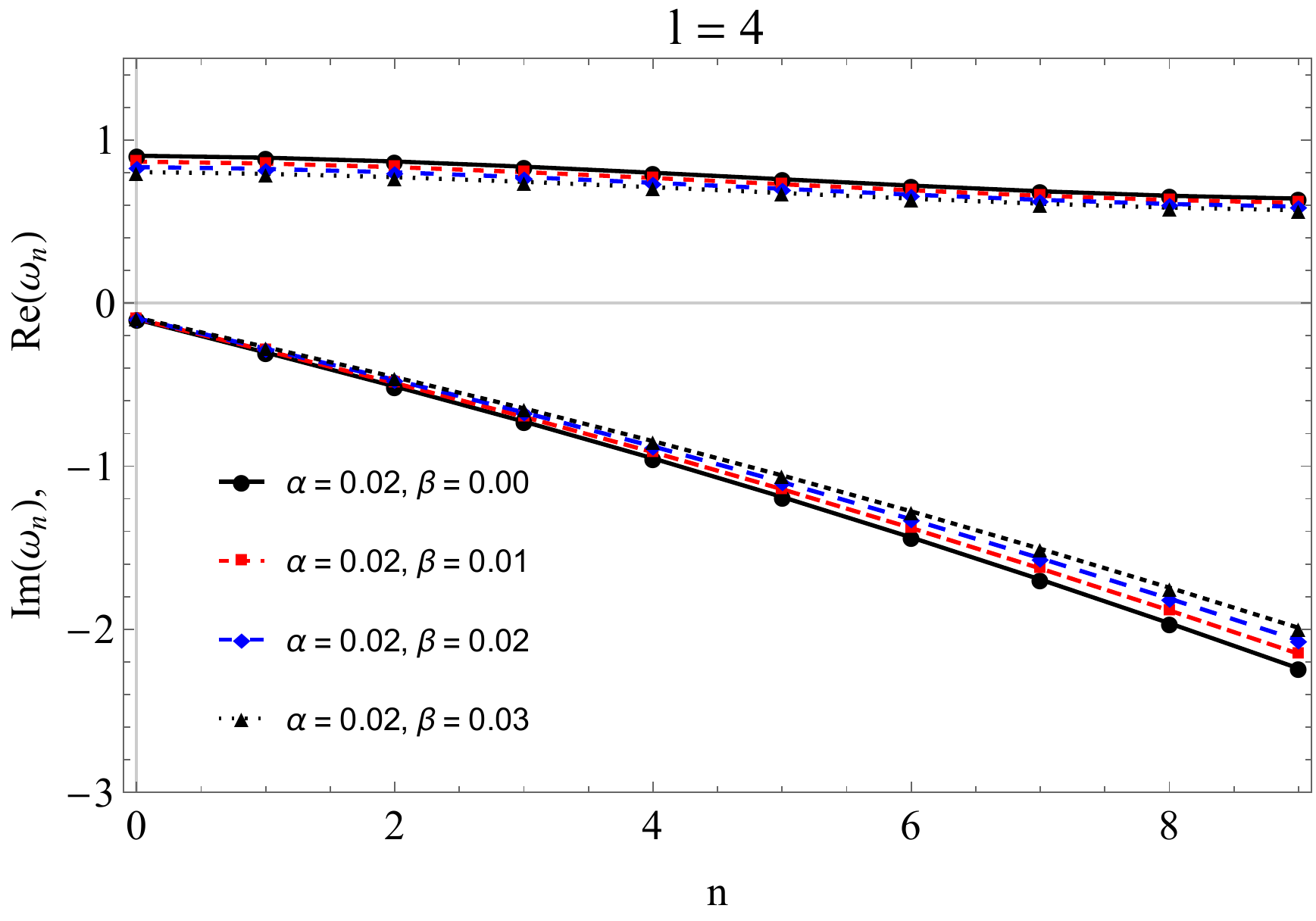}\label{figQN2l4}}
 \qquad
 \subfigure[]{\includegraphics[scale=0.30]{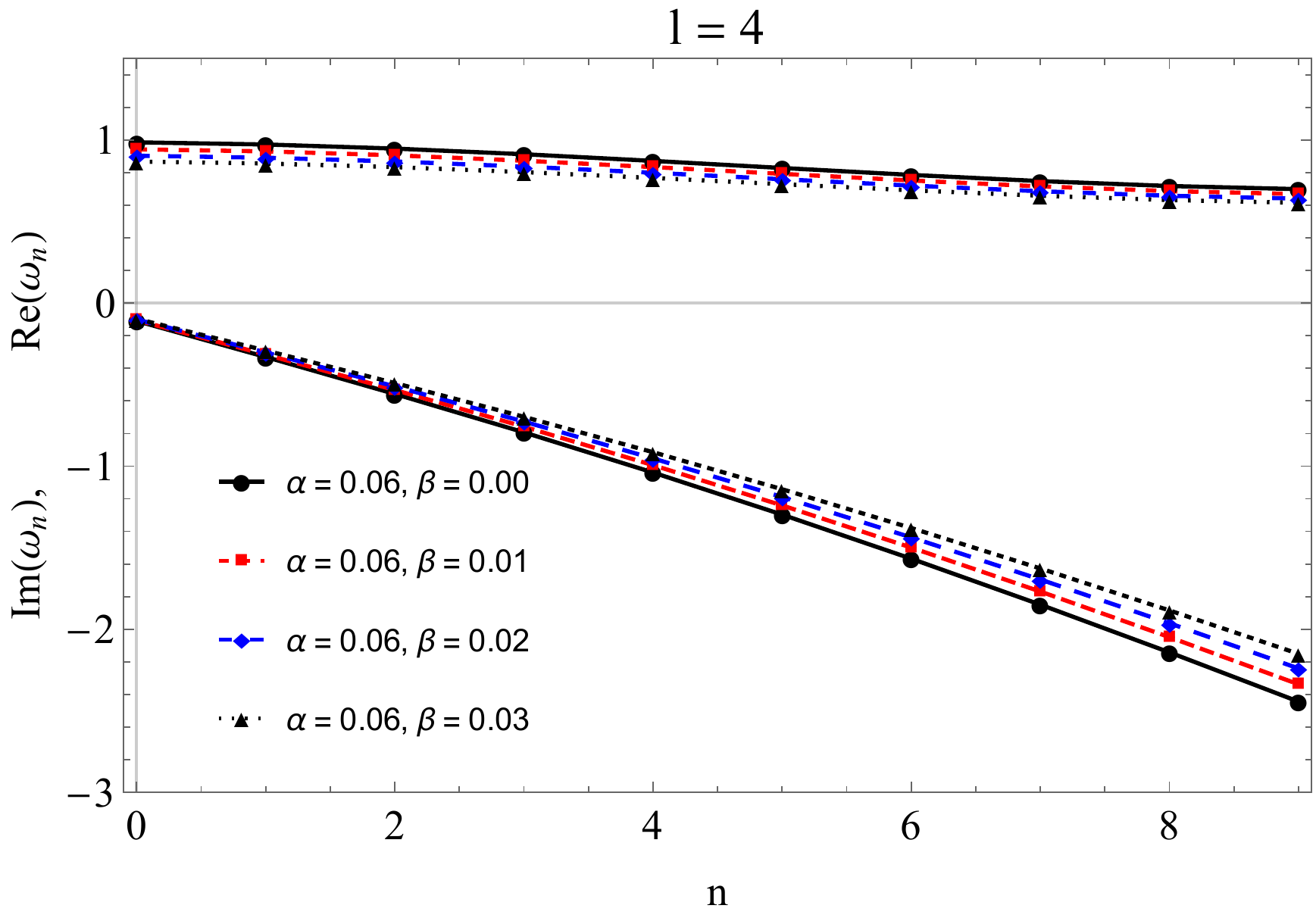}\label{figQN3l4}}
 \\
 \caption{\footnotesize{Quasinormal frequencies, real and imaginary parts versus normal $ n $ modes. The graphics are divided into real (top) and imaginary (bottom). We have a top to bottom distribution in ascending order for the number of multipoles from $ l = 1 $ to $ l = 4 $, while from left to right we have the variation of the parameter $ \alpha $ from 0, 0.02 to 0.06. In all graphs we assume the following values: $\beta = 0, 0.01, 0.02, 0.03 $. Note that when we change the values of $ \alpha $ (left to right) we have a very subtle variation in the curves so that it is more perceptible for $ n> 5 $.}} 
 \label{fig_Qn}
\end{figure}

\begin{figure}[!htb]
 \centering
 \subfigure[]{\includegraphics[scale=0.45]{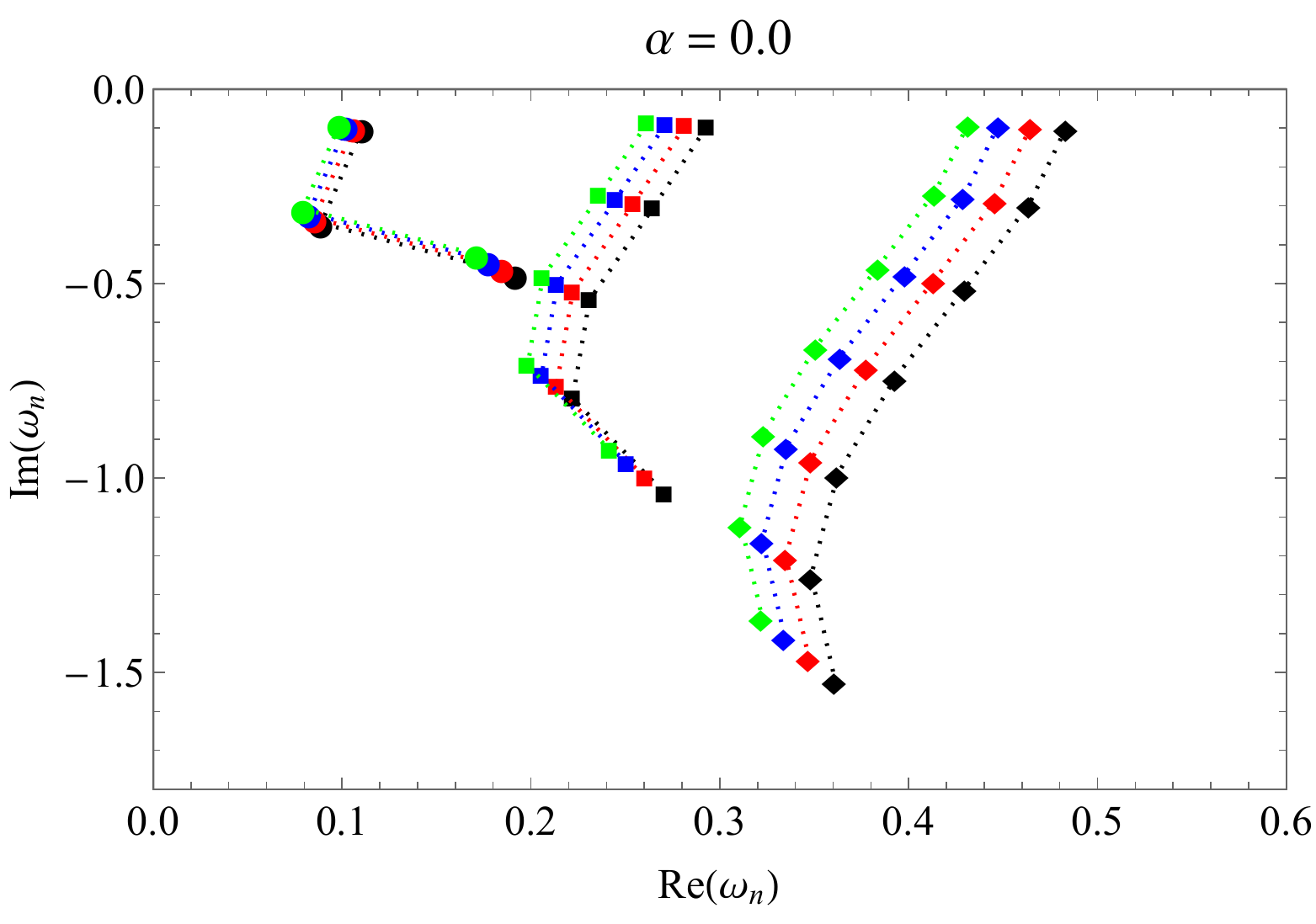}\label{plancomplex1}}
 \qquad
 \subfigure[]{\includegraphics[scale=0.45]{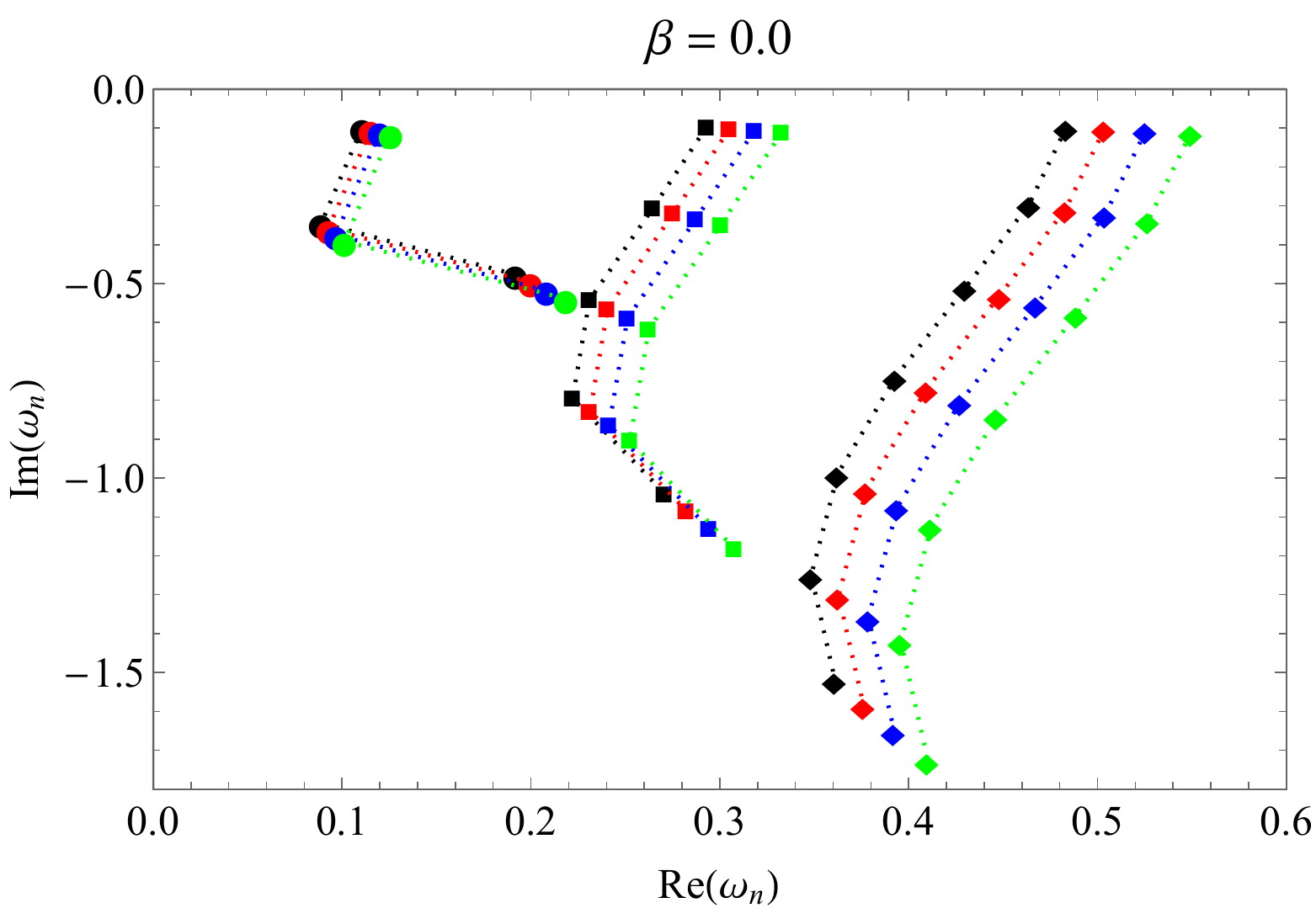}\label{plancomplex2}}
\caption{\footnotesize{Quasinormal modes in a complex plane for the following modes: $ l = 0 $ (circle), $ l = 1 $ (square), $ l = 2 $ (diamond). In (a) we consider only the quadratic part by making $ \alpha = 0 $ 
and $ \beta\geq 0 $: $ \beta = 0 $ (black), $ \beta = 0.01 $, (red) $ \beta = 0.02 $ (blue ) and $ \beta = $ 0.03 (green). In (b) we consider the linear part only, that is, $ \beta = 0 $ and $ \alpha \geq 0$: $ \alpha = 0 $ (black), $ \alpha = 0.02 $, (red) $ \alpha = 0.04 $ (blue) and $ \alpha = 0.06 $ (green).}}
\label{plancomplex}
\end{figure}

\section{Null geodesic and Shadow}
\label{ng-sh}

In this section, we aim to study the shadow radius of the Schwarzschild black hole with quantum corrections due to GUP. For this purpose, we will use the null geodesic equations.
So, starting with the Lagrangian
\begin{equation}
\mathcal{L} \equiv \dfrac{1}{2}g_{\mu\nu}\dot{x}^{\mu}\dot{x}^{\nu},
\end{equation} 
and applying the background metric~\eqref{metrsd}, we can write
\begin{equation}
2\mathcal{L} = {\cal F}(r)\dot{t}^{2} - \dfrac{\dot{r}^{2}}{{\cal F}(r)} - r^{2}\left( \dot{\theta}^{2} + \sin^{2}\theta \dot{\phi}^{2} \right),
\label{elidot}
\end{equation} 
where the ``." is the derivative with respect to the affine parameter.

We are interested in the path of a ray of light in the described metric, which is spherically symmetric, 
so if we analyze in a plane, any ray of light that begins with a certain angle $ \theta $ must remain with the same angle. We will then consider an equatorial plane by setting the angle $ \theta $ to $ \pi/2 $.
Thus, two equations are enough to describe the movement of a beam of light. We can put together a system with these equations that give rise to two geodesic motion constants $ E $ and $ L $, which correspond to energy and angular momentum respectively:
\begin{eqnarray}
E = {\cal F}(r)\dot{t}, \qquad L = r^{2}\dot{\phi}.
\label{eqEL}
\end{eqnarray}
For the case of null geodesics $g_{\mu\nu}\dot{x}^{\mu}\dot{x}^{\nu} = 0$, we obtain 
\begin{equation}
\dot{r}^{2} = E^{2} - {\cal F}(r)\dfrac{L^{2}}{r^{2}}.
\label{eqEner}
\end{equation}
From the above equation, we can write
\begin{equation}
\dot{r}^{2} +  U(r) = 0,
\label{eqVnull}
\end{equation}
where 
\begin{eqnarray}
U(r) = {\cal F}(r)\dfrac{L^{2}}{r^{2}} - E^{2}.
\end{eqnarray}
Now, applying the following conditions:
\begin{equation}
U(r_{c}) = 0, \qquad \dfrac{dU(r)}{dr}\Bigr\rvert_{r \rightarrow r_{c}} = 0,
\end{equation}
we find the critical radius $ r_ {c} $ and the critical impact parameter $ b_ {c} $, 
which are given respectively by 
\begin{eqnarray}
r_{c} &=& 3M\left( 1 - \dfrac{2\alpha}{M} + \dfrac{4\beta}{M^2}\right), \\
b_{c} &=& \dfrac{r_{c}}{\sqrt{{\cal F}(r_{c})}} = 3\sqrt{3}M\left( 1 - \dfrac{2\alpha}{M} + \dfrac{4\beta}{M^2}\right).
\label{paramcrit}
\end{eqnarray} 
Note that, assuming $ \alpha = 0 $ and $ \beta = 0 $ we return to the Schwarzschild case, 
where $ r_{c} = 3M $ and $ b_{c} = 3 \sqrt{3} M $.
Now, to determine the size of the shadow of a black hole, we introduce the celestial coordinates given by
\begin{eqnarray}
\xi &=& \lim\limits_{r_{o} \to \infty}\left[- r_{o}^{2} \sin\theta_{o}\dfrac{d\phi}{dr}\Bigr\rvert_{\theta = \theta_{o}} \right] ,\label{coodcelesta}\\
\eta &=& \lim\limits_{r_{o} \to \infty}\left[r_{o}^{2} \dfrac{d\theta}{dr}\Bigr\rvert_{\theta = \theta_{o}}\right] ,
\label{coodcelestb}
\end{eqnarray}
where $\left(r_{o}, \theta_{o}\right)$ is the observer's position at infinity. 
In this case, for an observer in the equatorial plane, that is in $ \theta_o = \pi/2 $, we have the following relation
\begin{equation}
R_{s} \equiv \sqrt{\xi^{2} + \eta^{2}} = b_{c}=3\sqrt{3}M -6\sqrt{3}\alpha + \frac{12\sqrt{3}\beta}{M}. 
\label{rshw}
\end{equation}
We can show that the shadow radius $R_{s}$ can be related to the real part of the quasinormal frequencies  in the eikonal limit as done by several authors~\cite{stefanov2010connection,jusufi2020quasinormal,cuadros2020analytical,Moura:2021eln,Liu2021}. One way to obtain this is to relate the quasinormal modes to the geodesic study as done by Cardoso {\it et al}~\cite{cardoso2009geodesic}. They showed that the real part of the quasinormal modes in the eikonal limit is related to the angular velocity for the last null circular orbit $\Omega_{c}$ and the imaginary part is related to the Lyapunov exponent $\lambda$ which determines the unstable timescale of the orbit:
\begin{eqnarray}
\label{VelAng}
\omega_{QNM} = \Omega_{c} l - i\left(n + 1/2\right)|\lambda|.
\end{eqnarray}
The angular velocity is given by $\Omega_{c} = \dot{\phi}/\dot{t}$, so by using \eqref{eqEL}  and \eqref{paramcrit} we obtain a relationship with the critical impact parameter 
\begin{eqnarray}
\Omega_{c} =  \dfrac{\mathcal{F}(r_{c})}{r_{c}^2}b_{c} = \dfrac{1}{b_{c}}.
\end{eqnarray}
This allows us to have, in the eikonal regime, a relationship between the quasinormal modes and the shadow radius of the black hole simply given by
 \begin{eqnarray}
\Re(\omega) = \lim_{l >> 1} \dfrac{l}{R_{s}}. 
\label{QNMandRs}
\end{eqnarray}
Although this relationship is valid for large values of $l$ in many cases it may not be valid in general, as shown by Konoplya and Stuchlik~\citep{konoplya2017eikonal} for Einstein-Lovelock theory.
In Fig.~\ref{figQNMandRs} we have the behavior of the shadow radius $\left(R_{s}\right)^{-1}$ as a function of $l$. The lines represent the values obtained by \eqref{rshw}, while the curves represent the values obtained by WKB approach using the parameters  $\alpha = 0.03$, $\beta = 0.01, 0.02, 0.03 $ and $n = 5$. We can clearly see that for large values of $l$ the curves (black, blue and red) fit the corresponding lines, which is the expected behavior according to Eq.~\eqref{QNMandRs}.

\begin{figure}[!htb]
 \centering
 \includegraphics[scale=0.4]{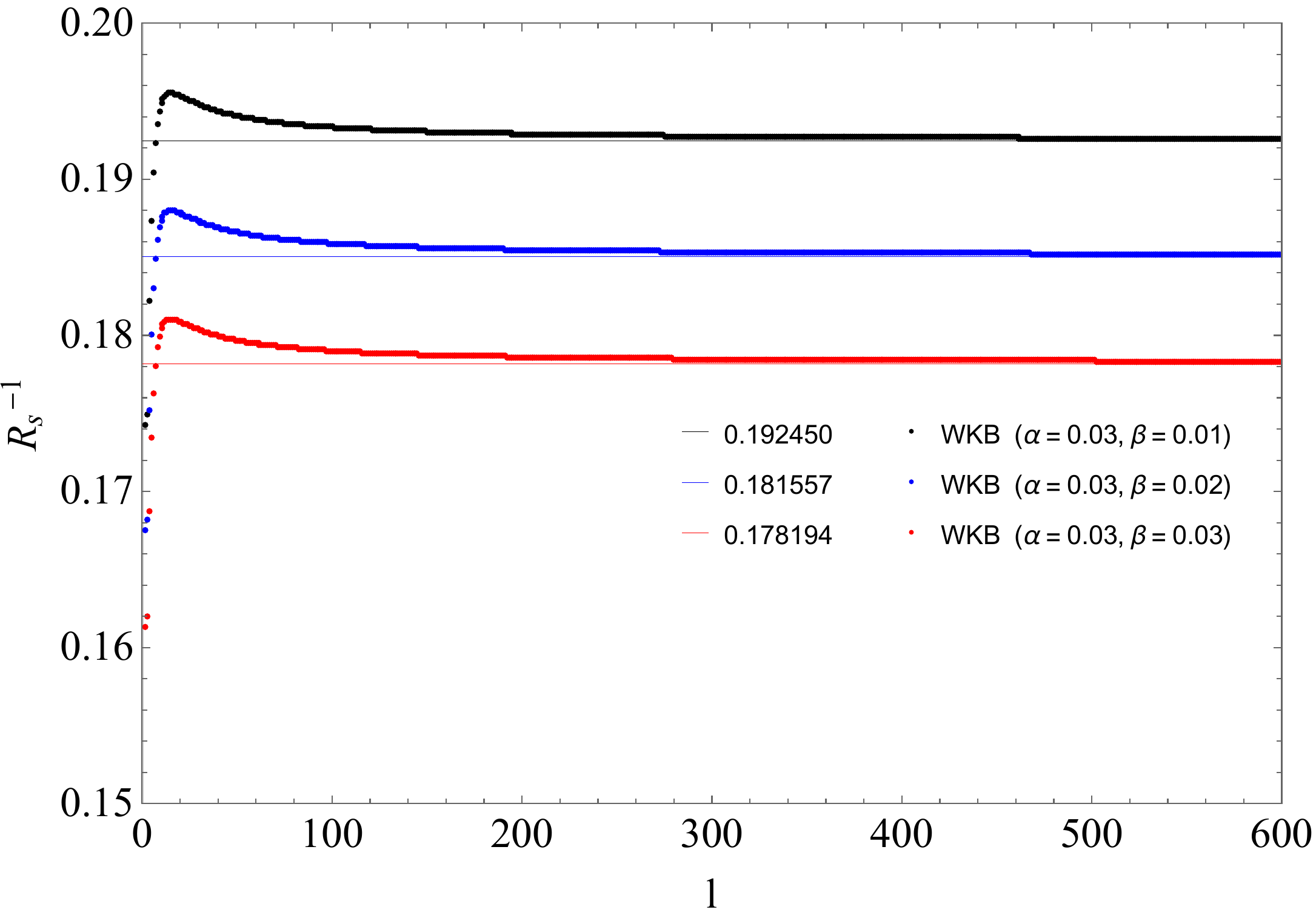}
 \caption{\footnotesize{The lines represent the values for $R_{s}^{-1} = 0.192450, 0.181557$ and $0.178194$, and the curves represent the values obtained by the WKB method for the corresponding parameters $\alpha$ and $\beta$. We assume $M = 1$ and $n = 5$ in the WKB approximation.}} 
  \label{figQNMandRs}
\end{figure}
In Fig.~\ref{shadow}, we see the shadows due to quantum corrections for a black hole with GUP. We show the behavior for different values of $ \alpha $ and $ \beta $. 
Note that, considering only the contribution of the quadratic GUP ($ \alpha=0 $ and $ \beta\neq 0 $), the shadow radius increases when we vary the parameter $ \beta $, as shown in Fig.~\ref{shaa0}. 
In Fig.~\ref{shaa006}, the increase in the shadow radius can also be observed when we set the parameter $\alpha$ and vary the parameter $ \beta $. The effect of linear GUP ($ \alpha\neq 0 $ and $ \beta=0 $)  is shown 
in Fig.~\ref{shab0}. In this case, the shadow radius is reduced when we vary $\alpha$.
In Fig.~\ref{shab003}, this effect can also be seen when we fix the value of $ \beta $ and increase the value of $\alpha$. 
In addition, we observe that for the quadratic contribution we just have that $r_{hgup} > r_h=2M$ and the critical radius of the photon sphere is greater than the radius $r_{hgup}$ ($ r_c > r_{hgup} $). 
Thus, for the shadow radius, our result shows that the quadratic part of the GUP generates an effect similar to that obtained by the contribution of dark matter in the vicinity of a black hole~\cite{Konoplya:2019sns}.

\begin{figure}[h!]
\centering
 \subfigure[]{\includegraphics[scale=0.35]{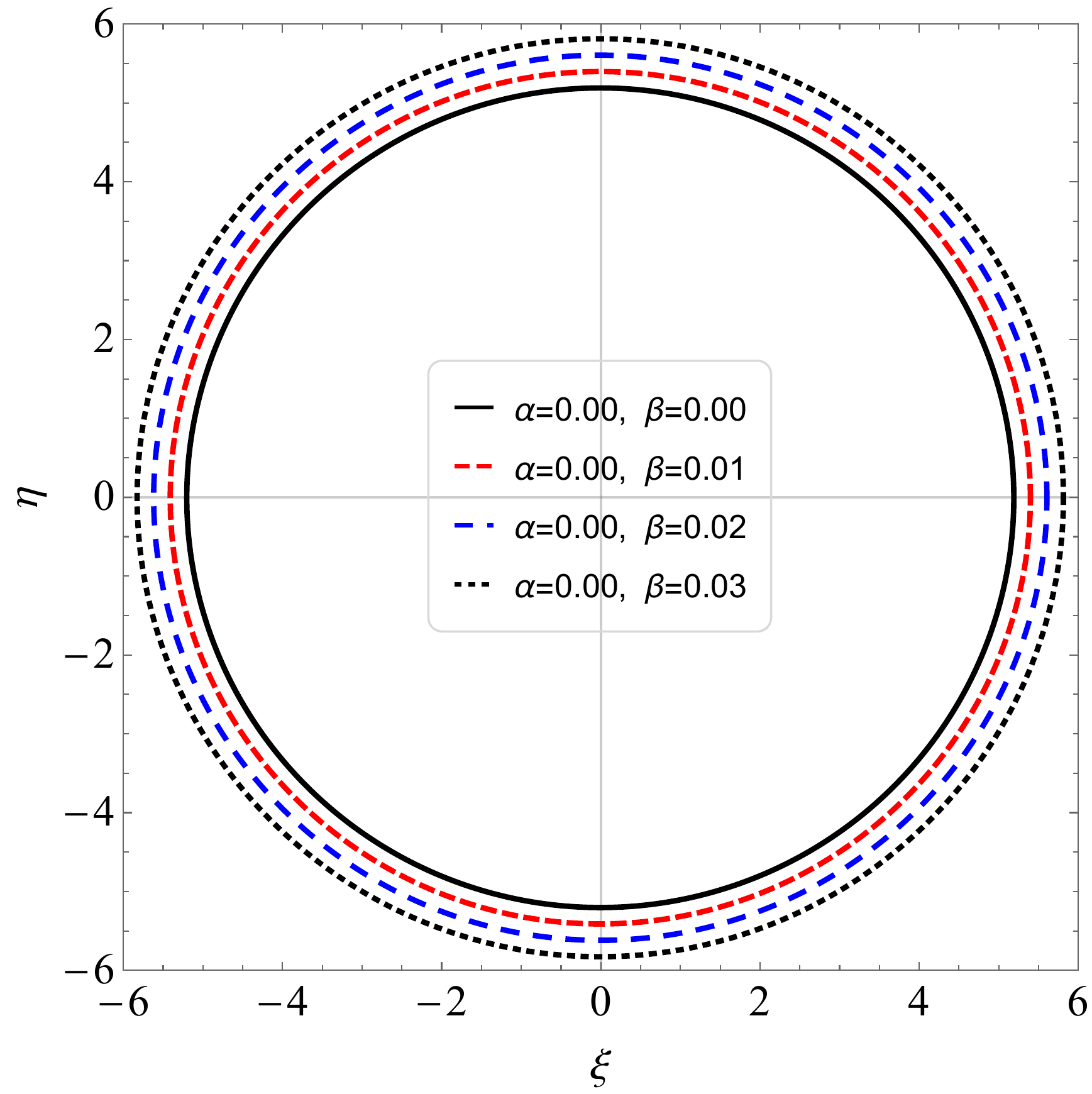}\label{shaa0}}
 \qquad
 \subfigure[]{\includegraphics[scale=0.35]{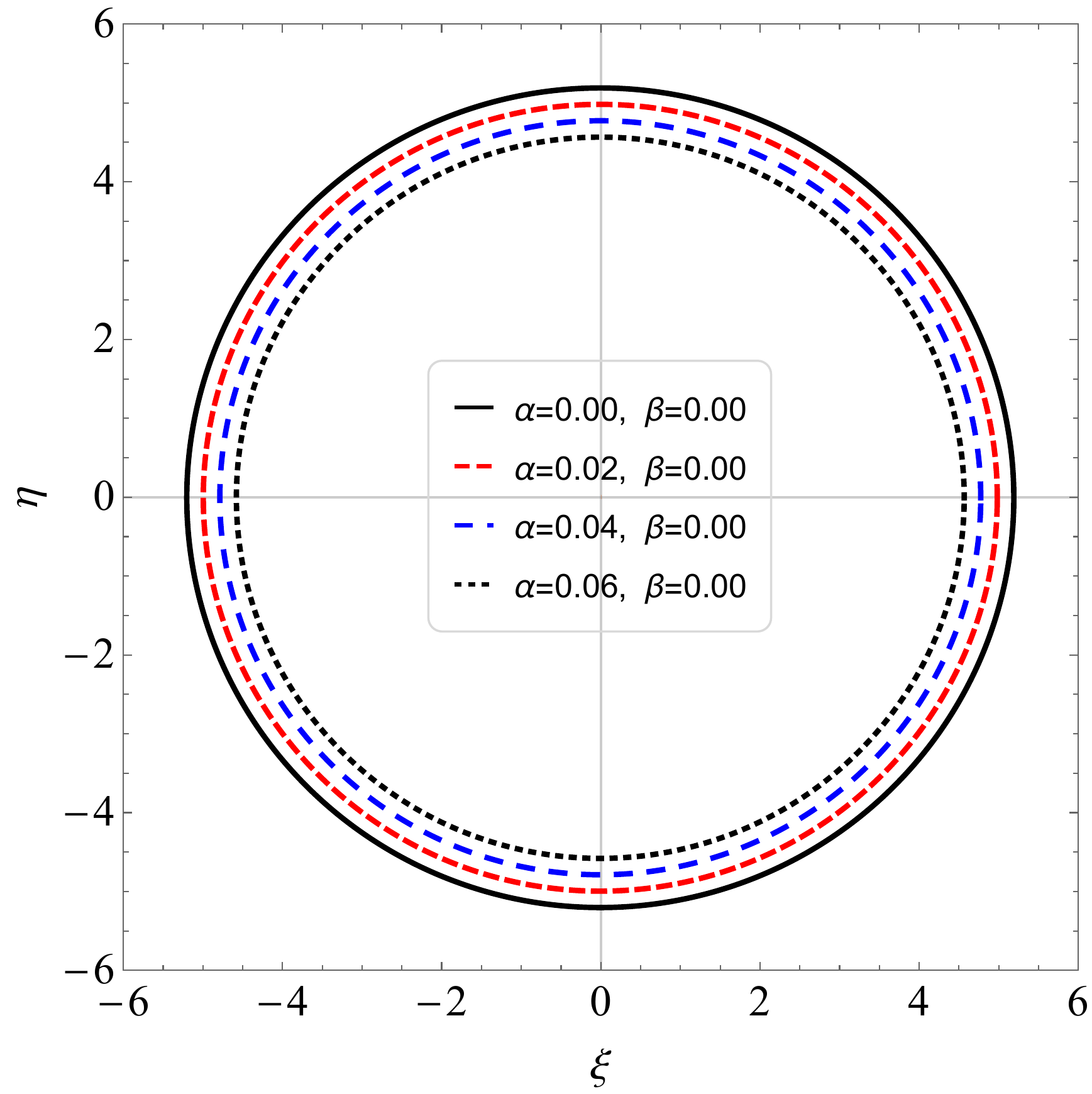}\label{shab0}}
 \qquad
 \subfigure[]{\includegraphics[scale=0.35]{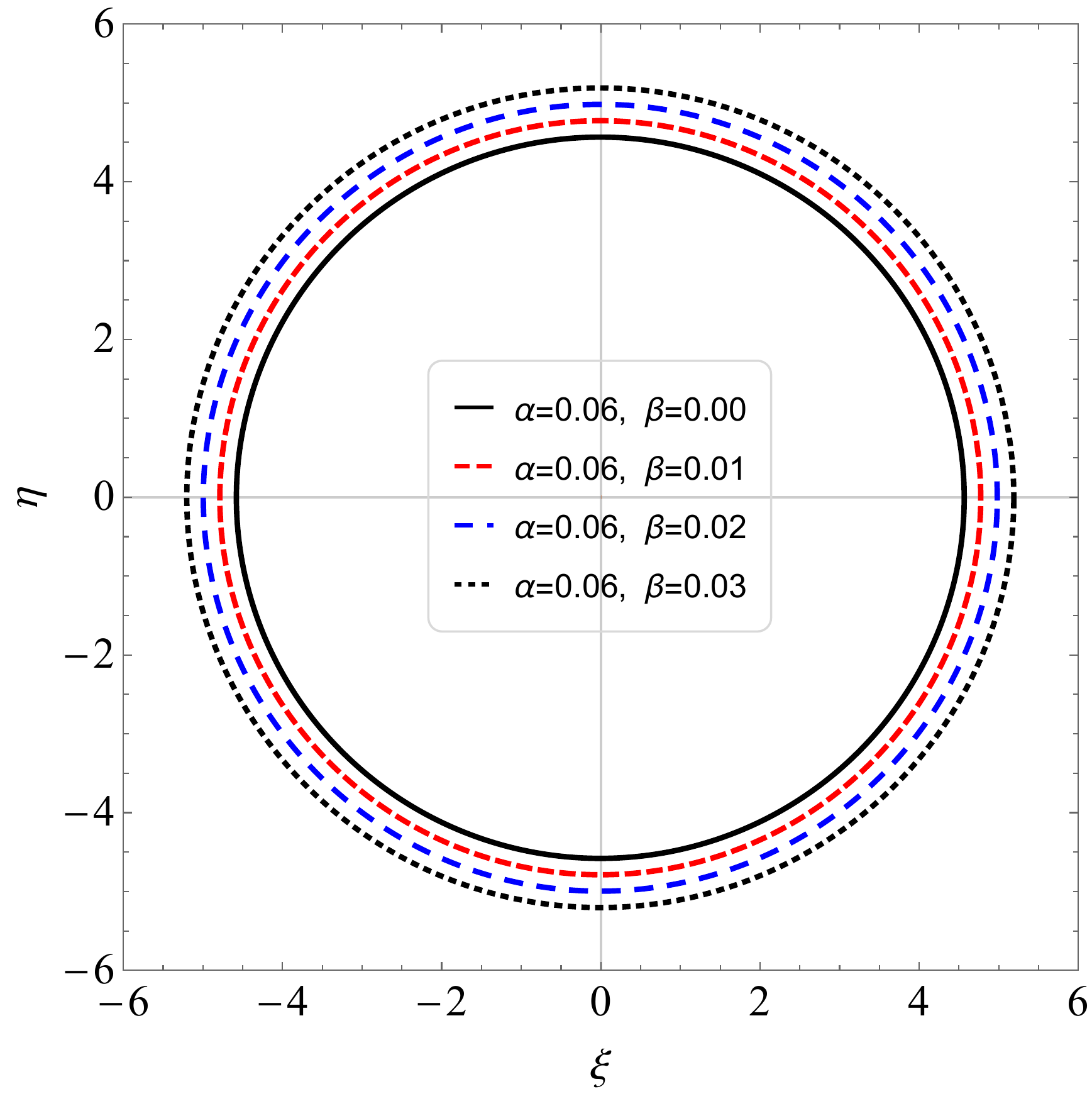}\label{shaa006}}
 \qquad
 \subfigure[]{\includegraphics[scale=0.35]{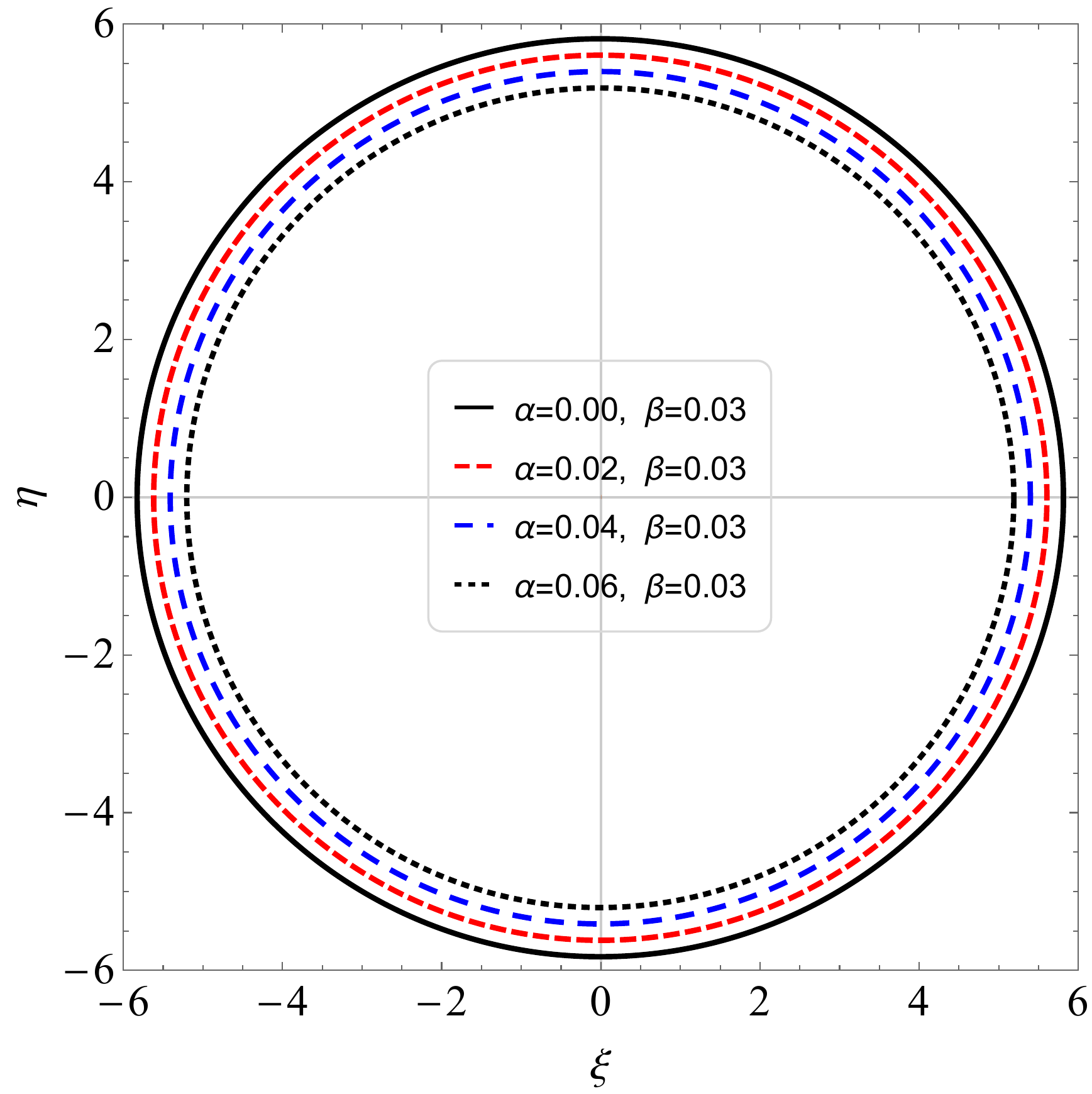}\label{shab003}}
 \caption{Each circle represents the shadow limit of the black hole with GUP, setting $ M = 1 $ for different values of $ \alpha $ and $ \beta $. }
 \label{shadow}
\end{figure} 

We can also, by taking the zero mass limit in the equation \eqref{rshw}, obtain a non-zero result given by the following expression
\begin{equation}
R_{s} \approx \dfrac{12\sqrt{3}\beta}{M}.
\label{shawM0}
\end{equation}
In Fig.~\ref {shadowM005}, we observe the behavior of the shadow radius for $ M = 0.05 $, which increases significantly with the addition of the parameter $ \beta $, while the change of values for the parameter 
$ \alpha $ does not have much influence on the results when we assume nonzero $ \beta $. This can be seen 
in \eqref{rshw}, and this effect is consistent with the results obtained in~\cite{Anacleto:2020lel}, which shows an increase in the absorption cross section to small $ M $. Here we have that the larger the critical radius, the larger the absorption region.
\begin{figure}[h!]
\centering
 \includegraphics[scale=0.35]{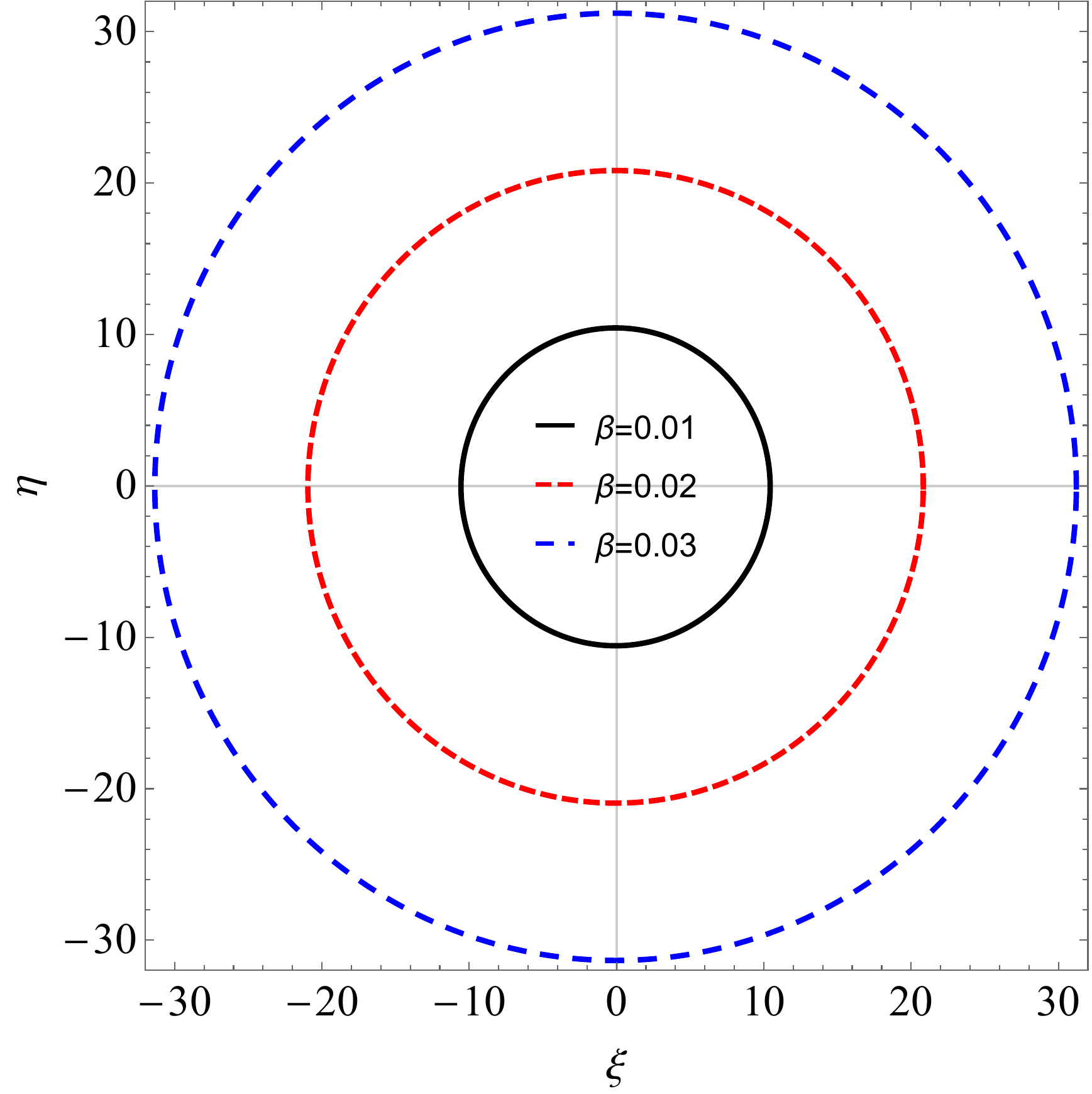}\label{shaM005}
 \caption{Circles for the shadows radius at the limit of small masses, by using the following values of parameters: $ M = 0.05 $, $ \alpha = 0 $ and $ \beta = 0.01, 0.02, 0.03 $.}
 \label{shadowM005}
\end{figure}

\section{Conclusions}
\label{conclu}

In this work we have computed the quasinormal modes of the quantum-corrected Schwarzschild black hole implemented by the GUP. 
By considering a Schroedinger-like equation, we have investigated scalar disturbances by using the sixth-order WKB approximation method. 
The quasinormal frequencies for a stable black hole are given by $ \omega=\omega_R -i\omega_I $, 
where $ \omega_R $ represents the frequency of the oscillation and $ \omega_I $ provides the damping time scale with positive $ \omega_I $ frequencies. 
According to our analyses the quasinormal frequencies follow precisely this behavior, then we have shown that all the obtained modes are stable.
We also have performed a numerical analysis whose results have confirmed the WKB approach.
In addition, we have graphically shown the real and imaginary parts of the quasinormal modes on the complex plane for three multipole families $l = 0, 1, 2$. By comparing with the Schwarzschild case, the curves associated with the quadratic part of the GUP ($\alpha=0$) move to the left with the increase of the parameter $\beta$, while for the linear part ($\beta=0$), the curves move to the right, when we increase the parameter $\alpha$. Notice that whereas in the former case one approaches smaller frequencies in the latter case it goes towards large frequencies. This is indeed in agreement with the effect of the linear and quadratic contribution of the GUP on the calculation of the shadow radius. We show that the quadratic part of the GUP has an effect of increasing the shadow radius when we change the parameter  $\beta$, while for the linear part the shadow radius is reduced when we change the parameter $\alpha$. As a consequence these two effects are in accord with the relationship of the real part of quasinormal modes with inverse of shadow radius, which means that in the present case the connection between the null geodesics and
quasinormal modes is not violated as it happens in the Einstein-Lovelock theory.
Finally, we also show that by keeping the quadratic part of the GUP the shadow radius does not vanish even at the limit of very small black hole mass. {Further investigations concerning the above  issues by considering, say,  charged scalar  particles in spacetime dimensions other than four such as in \cite{Gonzalez:2017zdz} should be addressed elsewhere.}

\acknowledgments
We would like to thank CNPq, CAPES and CNPq/PRONEX/FAPESQ-PB (Grant nos. 165/2018 and 015/2019),
for partial financial support. MAA, FAB and EP acknowledge support from CNPq (Grant nos. 306962/2018-7 and 433980/2018-4, 312104/2018-9, 304852/2017-1).

\end{document}